\documentclass[11pt,a4paper]{article}
\usepackage{amsfonts}
\usepackage{amssymb}
\usepackage[centertags]{amsmath}
\usepackage{graphicx}
\usepackage{booktabs}
\usepackage{theorem}
\usepackage{array}
\usepackage{colordvi}
\usepackage{ulem}
\usepackage[small,bf]{caption}
\usepackage{cite}
\usepackage{color}
\usepackage{subfigure}
\newcommand{\meff}{\mbox{$\left|  \langle \!\,  m  \,\!  \rangle \right| $}}
\newcommand{\betabeta}{\mbox{$(\beta \beta)_{0 \nu}  $}}
\def\ltap{\ \raisebox{-.4ex}{\rlap{$\sim$}} \raisebox{.4ex}{$<$}\ }

\allowdisplaybreaks[1]

0.4in \textheight 8.8in \numberwithin{equation}{section}

\newcommand{\be}{\begin{equation}}
\newcommand{\ee}{\end{equation}}
\newcommand{\ba}{\begin{eqnarray}}
\newcommand{\ea}{\end{eqnarray}}

\newcommand{\ti}[1]{\ensuremath{\tilde{#1}}} % \ti \psi
\newcommand{\si}{\textbf{1}}
\newcommand{\db}{\textbf{2}}
\newcommand{\tp}{\textbf{3}}
\newcommand{\TBM}{U_{\mbox{\tiny TBM}}}

\DeclareMathOperator{\diag}{Diag}
\DeclareMathOperator{\ci}{\text{i}}

\definecolor{pink}{rgb}{0.8, 0, 0.7}

\makeatletter

\newcommand{\Rmnum}[1]{\expandafter\@slowromancap\romannumeral #1@}
\makeatother

\begin{document}

\begin{titlepage}

\vspace*{-15mm}
\begin{flushright}
SISSA 12/2012/EP
\end{flushright}
\vspace*{0.7cm}

\begin{center}
{ \bf\LARGE A SUSY $\boldsymbol{SU(5) \times T^{\prime}}$ Unified Model of Flavour \\
with large $\boldsymbol{\theta_{13}}$}
\\[8mm]
Aurora Meroni$^{\, a}$,
\footnote{E-mail: \texttt{aurora.meroni@sissa.it}},
S. T. Petcov$^{\, a, b}$,
\footnote{Also at:
 Institute of Nuclear Research and Nuclear Energy,
  Bulgarian Academy of Sciences, 1784 Sofia, Bulgaria.},
Martin~Spinrath$^{\, a}$
\footnote{E-mail: \texttt{spinrath@sissa.it}},
\\[1mm]
\end{center}
\vspace*{0.50cm}
\centerline{$^{a}$ \it
SISSA/ISAS and INFN,}
\centerline{\it
Via Bonomea 265, I-34136 Trieste, Italy }
\vspace*{0.2cm}
\centerline{$^{b}$ \it
Kavli IPMU, University of Tokyo, Tokyo, Japan}
\vspace*{1.20cm}

\begin{abstract}
\noindent We present a SUSY $SU(5) \times T^{\prime}$ unified
flavour model with type I see-saw mechanism of neutrino mass
generation, which predicts the reactor neutrino angle to be
$\theta_{13} \approx 0.14$  close to  the recent results from the
Daya Bay and RENO experiments. The model predicts also values of the
solar and atmospheric neutrino mixing angles, which are compatible
with the existing data. The $T^{\prime}$ breaking leads to
tri-bimaximal mixing in the neutrino sector, which is perturbed by
sizeable corrections from the charged lepton sector. The model
exhibits geometrical CP violation, where all complex phases have
their origin from the complex Clebsch--Gordan coefficients of
$T^{\prime}$. The values of the Dirac and Majorana CP violating
phases are predicted. For the Dirac phase in the standard
parametrisation of the neutrino mixing matrix we get a value close
to $90^{\circ}$: $\delta \cong \pi/2 - 0.45 \, \theta^c \cong
84.3^{\circ}$, $\theta^c$ being the Cabibbo angle. The neutrino mass
spectrum can be with normal ordering (2 cases) or inverted ordering.
In each case the values of the three light neutrino masses are
predicted with relatively small uncertainties, which allows to get
also unambiguous predictions for the neutrino-less double
beta decay effective Majorana mass.
\end{abstract}

\end{titlepage}
\setcounter{footnote}{0}

\section{Introduction}
  Understanding the origin of the patterns of neutrino
masses and mixing, emerging from the neutrino oscillation,
$^3H$ $\beta-$decay, etc. data is one of the most
challenging problems in neutrino physics.
It is part of the more general fundamental problem
in particle physics of understanding the origins of
flavour, i.e., of the patterns
of the quark, charged lepton and neutrino masses
and of the quark and lepton mixing.

   At present we have compelling evidence for
existence of mixing of three light massive neutrinos $\nu_i$,
$i=1,2,3$, in the weak charged lepton current (see, e.g.,
\cite{PDG10}). The masses $m_i$ of the three light neutrinos $\nu_i$
do not exceed approximately 1 eV, $m_i \ltap 1$ eV, i.e., they are
much smaller than the masses of the charged leptons and quarks. The
three light neutrino mixing is described (to a good approximation)
by the Pontecorvo, Maki, Nakagawa, Sakata (PMNS) $3\times 3$ unitary
mixing matrix, $U_{\rm PMNS}$. In the widely used standard
parametrisation \cite{PDG10},  $U_{\rm PMNS}$ is expressed in terms
of the solar, atmospheric and reactor neutrino mixing angles
$\theta_{12}$,  $\theta_{23}$ and $\theta_{13}$, respectively, and
one Dirac - $\delta$, and two Majorana \cite{BHP80}  - 
$\beta_{1}$ and $\beta_{2}$, CP violating phases:
%%%%%%%%%%%%%%%%%%%%%%%%%%%
\be
 U_{\text{PMNS}} \equiv U = V(\theta_{12},\theta_{23},\theta_{13},\delta)\,
Q(\beta_1,\beta_2)\,, \label{eq:UPMNS} \ee
%%%%%%%%%%%%%%%%%%%%%%%%%%%
%
where
%%%%%%%%%%%%%%%%%%%%%%%%%%%%%%%
\be
V = \left(
     \begin{array}{ccc}
       1 & 0 & 0 \\
       0 & c_{23} & s_{23} \\
       0 & -s_{23} & c_{23} \\
     \end{array}
   \right)\left(
            \begin{array}{ccc}
              c_{13} & 0 & s_{13}e^{-\ci \delta} \\
              0 & 1 & 0 \\
              -s_{13}e^{\ci \delta} & 0 & c_{13} \\
            \end{array}
          \right)\left(
                   \begin{array}{ccc}
                     c_{12} & s_{12} & 0 \\
                     -s_{12} & c_{12} & 0 \\
                     0 & 0 & 1 \\
                   \end{array}
                 \right)\,,
\label{eq:V}
\ee
%%%%%%%%%%%%%%%%%%%%%%%%%%%%%%%%%%%
%
and we have used the standard notation $c_{ij} \equiv
\cos\theta_{ij}$, $s_{ij} \equiv \sin\theta_{ij}$ and
\footnote{
This parametrization differs from the
standard one. We use it for ''technical'' reasons
related to the fitting code we will employ. Obviously, 
the standard one can be obtained as
$\diag(1, e^{\ci \alpha_{21}}, e^{\ci \alpha_{31}})= e^{\ci
\beta_1/2}Q$, with $\alpha_{21} = \beta_1 - \beta_2$ and
$\alpha_{31} = \beta_1$~.}
%%%%%%%%%%%%%%%%%%%%%%%%%%%%%%%%%
\be Q = \diag(e^{-\ci \beta_1/2}, e^{-\ci \beta_2/2},1)\,. \label{Q}
\ee
%%%%%%%%%%%%%%%%%%%%%%%%%%%%%%%

 The neutrino oscillation data, accumulated over many years, allowed
to determine the parameters which drive the solar and atmospheric
neutrino oscillations, $\Delta m^{2}_{\odot}\equiv \Delta
m^{2}_{21}$, $\theta_{12}$ and $|\Delta m_A^2| \equiv |\Delta
m^{2}_{31}| \cong |\Delta m^{2}_{32}|$, $\theta_{23}$, with a rather
high precision (see, e.g., \cite{PDG10}). Furthermore, there were
spectacular developments in the last year in what concerns the angle
$\theta_{13}$. In June of 2011 the T2K collaboration  reported
\cite{Abe:2011sj} evidence at $2.5\sigma$ for a non-zero value of
$\theta_{13}$. Subsequently the MINOS \cite{MINOS240611} and Double
Chooz \cite{DChooz2011} collaborations also reported evidence for
$\theta_{13}\neq 0$, although with a smaller statistical
significance. Global analysis of the neutrino oscillation data,
including the data from the T2K and MINOS experiments, performed in
\cite{Fogli:2011qn}, showed  that actually $\sin\theta_{13}\neq 0$
at $\geq 3\sigma$. In March of 2012 the first data of the Daya Bay
reactor antineutrino experiment on $\theta_{13}$ were published
\cite{An:2012eh}. The value of $\sin^22\theta_{13}$ was measured
with a rather high precision and was found to be different from zero
at $5.2\sigma$:
%%%%%%%%%%%%%%%%%%%%%%%%%
\begin{equation}
 \sin^22\theta_{13} = 0.092 \pm 0.016 \pm 0.005\,,~~
 0.04 \leq \sin^22\theta_{13} \leq 0.14\,,~3\sigma\,,
\label{DBayth13}
\end{equation}
%%%%%%%%%%%%%%%%%%%%%%%%
%
where we have given also the $3\sigma$ interval
of allowed values of  $\sin^22\theta_{13}$.
Subsequently, the RENO experiment
reported a $4.9\sigma$ evidence
for a non-zero value of $\theta_{13}$ \cite{RENO0412},
compatibe with the Day Bay result:
%%%%%%%%%%%%%%%%%%%%%%%%%
\begin{equation}
 \sin^22\theta_{13} = 0.113 \pm 0.013 \pm 0.019\,.
\label{RENOth13}
\end{equation}
%%%%%%%%%%%%%%%%%%%%%%%%
%
 The results on $\theta_{13}$ described above
will have far reaching implications
for the program of future research in neutrino physics
(see, e.g., \cite{MMexTSchwth13}).

  A recent global analysis of the current
neutrino oscillation data, in which
the Daya Bay and RENO results on $\theta_{13}$ are also included, was
published \cite{Tortola:2012te}.
In Table \ref{tab:tabdata-1205}  we show the best fit values and
the 99.73\% CL allowed ranges of
$\Delta m^2_{21}$, $\sin^2\theta_{12}$,
$|\Delta m^2_{31(32)}|$ , $\sin^2\theta_{23}$ and $\sin^2\theta_{13}$,
found in  \cite{Tortola:2012te}.
%%%%%%%%%%%%%%%%%%%%%%%%%%%%%%%%%%%%%%%%

%%%%%%%%%%%%%%%%%%%%%%%%%%%%%%%%%%%%%%%%
\begin{table}
\centering
\renewcommand{\arraystretch}{1.1}
\begin{tabular}{lcc}
\toprule
 Parameter  &  best-fit ($\pm 1\sigma$) & 3$\sigma$ \\ \midrule
 $\Delta m^{2}_{\odot} \; [10^{-5}\text{ eV}^2]$   & 7.62$\pm 0.19$ &
               7.12 - 8.20 \\
$ |\Delta m^{2}_{A}| \; [10^{-3}\text{ eV}^2]$ & 2.53$^{+0.08}_{-0.10}$  &
           2.26 - 2.77\\
& -(2.40$^{+0.10}_{-0.07}$) &  -(2.15-2.68)\\
 $\sin^2\theta_{12}$  & 0.320$^{+0.015}_{-0.017}$
            & 0.27 - 0.37\\
$\sin^2\theta_{23}$  & 0.49$^{+0.08}_{-0.02}$ &  0.39-0.64 \\
            &  0.53$^{+0.05}_{-0.07}$  & \\
$\sin^2\theta_{13}$  &  0.026$^{+0.003}_{-0.004}$  & 0.015-0.036\\
 & 0.027$^{+0.003}_{-0.004}$ & 0.016-0.037 \\
\bottomrule
\end{tabular}
\caption{ \label{tab:tabdata-1205} The best-fit values
and $3\sigma$ allowed ranges of the 3-neutrino oscillation
parameters derived from a global fit of the current
neutrino oscillation data, including  the Daya Bay and RENO
results (from \cite{Tortola:2012te}).
These values
are obtained using the ``new''\cite{Mention11}
reactor $\bar{\nu}_e$ fluxes. If two values are given the first one corresponds to normal hierarchy
and the second one to inverted hierarchy.
}
\end{table}
%%%%%%%%%%%%%%%%%%%%%%%%%%%%%%%%%%%%%%%%

  Stimulated by the fact that all three angles in the PMNS matrix are
determined with a relatively high precision, we report in the
present article an attempt to construct a unified model of flavour,
which describes correctly the quark and charged lepton masses, the
mixing and CP violation in the quark sector, the mixing in the
lepton sector, including the relatively large value of the angle
$\theta_{13}$, and provides predictions for the light neutrino
masses compatible with the existing relevant data and constraints.
The unified model of flavour we are proposing is
supersymmetric and is based
on $SU(5)$ as gauge group and $T^{\prime}$ as discrete family
symmetry.
 It includes three right-handed (RH)
neutrino fields $N_{lR}$, $l=e,\mu,\tau$, which possess a Majorana
mass term. The light neutrino masses are generated by the type I
see-saw mechanism \cite{seesaw} and are naturally small. The
corresponding Majorana mass term of the left-handed flavour neutrino
fields $\nu_{lL}(x)$, $l=e,\mu,\tau$, is diagonalised by a unitary
matrix which, up to a diagonal phase matrix, is of the tri-bimaximal
form~\cite{tri1}:
%%%%%%%%%%%%%%%%%%%%%%%%%%%%%%%%%%%%%%%%%%
\begin{equation}
\TBM = \left(\begin{array}{ccc}
\sqrt{2/3} & \sqrt{1/3} & 0 \\
-\sqrt{1/6} & \sqrt{1/3} & -\sqrt{1/2} \\
-\sqrt{1/6} & \sqrt{1/3} & \sqrt{1/2}
\end{array}\right) \;.
\label{TBMM}
\end{equation}
%%%%%%%%%%%%%%%%%%%%%%%%%%%%%%%%
%
In order to account for the current data
on the neutrino mixing, and more specifically,
for the fact that $\theta_{13} \neq 0$,
$\TBM$ has to be ``corrected''.
The requisite correction is provided
by the unitary matrix originating from the
diagonalisation of the charged lepton
mass matrix $M_{e}$ (for a general
discussion of such corrections see, e.g.,
\cite{FPR,HPR07,Marzocca:2011dh}).
Since the model is based on the $SU(5)$ GUT
symmetry, the charged lepton mass matrix $M_e$
is related to the down-quark mass matrix $M_d$.
As a consequence, in particular,
of the connection between  $M_{e}$ and  $M_d$,
the smallest angle in the neutrino
mixing matrix $\theta_{13}$,
is related to the Cabibbo angle $\theta^c$:
$ \sin^2\theta_{13}\cong C^2 (\sin^2\theta^c)/2
\cong (\sin^2\theta^c)/2.5$,
where  $C\cong 0.9$ is a constant determined
from the fit.

 The down-quark mass matrix $M_d$,
and the charged lepton mass matrix $M_e$,
by construction are neither diagonal nor
CP conserving. The matrix $M_e$ is the only source
of CP violation in the lepton sector.
Actually, the CP violation
predicted by the model in the quark and lepton sectors
is entirely geometrical in origin.
This aspect of the $SU(5)\times T'$ model we
propose is a consequence, in particular, of one of
the special properties of the group $T^{\prime}$
\footnote{There have been also $T^{\prime}$ models without a GUT embedding, e.g.\
\cite{Frampton:1994rk,Feruglio:2007uu, Ding:2008rj}.},
namely, that its group theoretical Clebsch-Gordan
(CG) coefficients are intrinsically complex~\cite{cg}.
The idea to use the complexity of the
Clebsch-Gordan (CG) coefficients of $T^{\prime}$
to generate the requisite CP violation
in the quark sector and a related
CP violation in the lepton sector was
pioneered in \cite{Chen:2009gf}.
For the class of models where the CP violation
is geometrical in origin,
it is essential to provide a solution to the vacuum
alignment problem for which all the flavon vevs
are real. In this paper we present
a solution of this problem for the models based
on the $SU(5)\times T^{\prime}$ symmetry.

 Let us note finally that a model of flavour
based on the symmetry group
 $SU(5)\times T^{\prime}$ was proposed, to our knowledge,
first in \cite{Chen:2007afa} and its
properties were further ellaborated in
\cite{Chen:2009gf} and \cite{Chen:2011vd}.
Although some generic features, as like
the connection between the reactor mixing
angle $\theta_{13}$ and the Cabibbo angle $\theta^c$,
which are based on the underlying $SU(5)$
symmetry, are present both in the model
constructed in \cite{Chen:2007afa,Chen:2009gf}
and in the model presented here,
the detailed structure and the quantitative
predictions of the two models are very different.
The quark, charged lepton, RH neutrino mass matrices and the
matrix of the neutrino Yukawa couplings have different
forms in the two models. This leads to considerable
differences in the predictions for various
observables.
In the quark sector, for instance,
the value of the CKM phase we find
is in much better agreement with experimental data.
More importantly, in the model proposed in
\cite{Chen:2007afa,Chen:2009gf},
the reactor mixing angle $\theta_{13}$ is predicted to have the
value $\sin\theta_{13} \cong \sin\theta^{c}/(3\sqrt{2}) \cong 0.016$,
which is ruled out by the current data on $\theta_{13}$.
In contrast, due to non-standard $SU(5)$ Clebsch--Gordan
relations between the down-type quark and the charged lepton
Yukawa couplings \cite{Antusch:2009gu,Marzocca:2011dh}, we get a realistic value for this angle.
Moreover, in the model we propose both neutrino mass spectra with
normal and inverted ordering  are possible, while  the model
developed in  \cite{Chen:2007afa,Chen:2009gf} admits only
neutrino mass spectrum with normal
ordering \cite{Chen:2011vd}.

   The paper is organized as follows. Section 2 is a brief 
overview of the considered model.
In  section 3 we discuss  the quark and charged lepton sector
including a $\chi^2$ fit to the experimental data.
Section 4 is completely devoted to the neutrino sector. 
There we describe  in detail
the predictions for the mixing parameters 
(including CP violating phases), the mass spectra and
observables such as the sum of the neutrino masses, 
the neutrinoless double beta ($\betabeta$-) decay
effective Majorana mass
and the rephasing invariant related to the Dirac phase
in the PMNS matrix, $J_{\text{CP}}$.
We summarize and conclude in section 5.
 In the Appendix we discuss the properties of 
the discrete group $T^{\prime}$,
the messenger sector which generates the effective 
operators for the Yukawa couplings,
and the superpotential, solving the flavon 
vacuum alignment problem.

\section{Matter, Higgs and Flavon Field
Content of the Model}

In this section we describe the matter, the Higgs and the
flavon content of our $SU(5)\times T^{\prime}$ unified model
of flavour. A rather large shaping symmetry,
$Z_{12} \times Z_8^3 \times Z_6^2 \times Z_4$, is needed to solve
the vacuum alignment issue and forbids unwanted terms and couplings
in the superpotential (specifically in the renormalisable one as
described in Appendix \ref{App:Messengers}, as well as in the
effective one after integrating out heavy messenger fields).  We
further impose an additional $U(1)_R$ symmetry, the continuous
generalisation of the
usual $R$-parity.
The messenger fields and auxiliary flavons used for the flavon
superpotential are discussed in the Appendix.

%%%%%%%%%%%%%%%%%%%%%%%%%%%%%
\begin{table}
\centering
\begin{tabular}{c cccc ccccccccc}
\toprule
& $T_3$ & $T_a$ & $\bar F$ & $N$ & $H_5^{(1)}$ & $H_5^{(2)}$ & $H_5^{(3)}$ & $\bar H_{5}^{(1)}$ & $\bar H_{5}^{(2)}$ & $\bar H_{5}^{(3)}$ & $\bar H_{5}^{\prime \prime}$  & $ H_{24}^{\prime \prime}$ & $ \tilde H_{24}^{\prime \prime}$  \\
\midrule $SU(5)$ & $\mathbf{10}$ & $\mathbf{10}$ & $\mathbf{\bar 5}$
& $\mathbf{1}$ & $\mathbf{5}$ & $\mathbf{5}$ & $\mathbf{5}$ & $\mathbf{\bar 5}$
& $\mathbf{\bar 5}$ & $\mathbf{\bar 5}$ & $\mathbf{\bar 5}$ & $\mathbf{24}$ & $\mathbf{24}$  \\
$T^\prime$ & $\mathbf{1}$ & $\mathbf{2}$ & $\mathbf{3}$  &
$\mathbf{3}$ & $\mathbf{1}$ & $\mathbf{1}$ & $\mathbf{1}$
& $\mathbf{1}$ & $\mathbf{1}$ & $\mathbf{1}$
& $\mathbf{1^{\prime \prime}}$ & $\mathbf{1^{\prime \prime}}$ & $\mathbf{1^{\prime \prime}}$ \\
$U(1)_R$ & 1 & 1 & 1 & 1 & 0 & 0 & 0 & 0 & 0 & 0 & 0 & 0 & 0 \\
$Z_{12}^u$ & 2& 11& 1& 9& 8& 8& 2& 9& 3& 6& 3& 0& 3\\
$Z_{8}^d$ & 4& 0& 2& 6& 0& 4& 0& 1& 4& 7& 7& 4& 2\\
$Z_{8}^\nu $ & 7& 6& 2& 0& 2& 6& 4& 1& 1& 5& 7& 4& 0\\
$Z_{8} $ &0& 5& 2& 2& 0& 0& 6& 0& 0& 6& 6& 4& 2\\
$Z_{6} $ & 5& 0& 1& 0& 2& 5& 2& 2& 0& 2& 2& 0& 0\\
$Z_{6}^{\prime} $ & 2& 3& 1& 0& 2& 5& 2& 5& 0& 2& 2& 0& 0\\
$Z_{4} $ & 3& 3& 0& 0& 2& 0& 2& 0& 1& 1& 0& 0& 1\\
\bottomrule
\end{tabular}
\caption{\label{tab:Matter+Higgs}
Matter and Higgs field content of the model including quantum numbers.}
\end{table}
%%%%%%%%%%%%%%%%%%%%%%%%%%%%%%%%%%%%

The model includes the three generations of matter fields in
the usual $\bar{\mathbf{5}}$ and $\mathbf{10}$,
representations of $SU(5)$, $\bar{F} = (d^c,L)_L$ and $T =
(q,u^c,e^c)_L$ and three heavy right-handed Majorana neutrino fields
$N$,
%set to be
singlets under $SU(5)$. The light active neutrino
masses are generated through the type I seesaw mechanism \cite{seesaw}.
Furthermore we introduce
a number of  copies of Higgs fields in the $\mathbf 5$ and
$\bar{ \mathbf{5}}$ representation of $SU(5)$ which contain as
linear combinations the two Higgs doublets of the MSSM. To get
realistic mass ratios between down-type quarks and charged leptons
\cite{Antusch:2009gu} and to get a large reactor mixing angle
\cite{Marzocca:2011dh} we have introduced Higgs fields in the
adjoint representation of $SU(5)$ which are as well responsible for
breaking the GUT group.

The matter and Higgs fields
including their transformation properties under all
imposed symmetries are summarised in Tab.\ \ref{tab:Matter+Higgs}.
Note that the right-handed neutrinos $N$ and the five-dimensional
matter representations are organised in $T^{\prime}$ triplets, while
the ten-plets are organised in a doublet and a singlet.
On the one hand this will give us tri-bimaximal mixing (TBM)
in the neutrino sector before considering corrections
from the charged lepton sector and on the other hand
the complex Clebsch--Gordan coefficients for the doublets
will give us CP violation in the quark and in the lepton
sector finally.

There are 13 flavons, which will give us the desired structure for
the Yukawa couplings that will be discussed
in the next section. First of all we have three triplets which
will develop vevs into two different directions in flavour space,
%%%%%%%%%%%%%%%%%%%%
\begin{equation}
\label{eq:3FlavonAlignment}
\langle \phi \rangle = \begin{pmatrix} 0 \\0 \\ 1 \end{pmatrix} \phi_{0}\;, \quad
\langle \ti\phi \rangle = \begin{pmatrix} 0 \\0 \\ 1 \end{pmatrix} \ti\phi_{0}\;, \quad
\langle \xi \rangle = \begin{pmatrix} 1 \\ 1 \\ 1 \end{pmatrix} \xi_{0} \;.
\end{equation}
%%%%%%%%%%%%%%%%%%%%%%%%%%%%%%
%
The first two flavons will be relevant for the quark and
the charged lepton sector and the third one couples only
to the neutrino sector.

Then we have introduced four complex $T^{\prime}$ doublets.
Notice that  this spinorial representations of the $T^{\prime}$
group are essential since, having complex Clebsch-Gordan coefficients
(see Appendix~\ref{App:Tprime}), it is responsible of the CP violation in both quark and
charged lepton sector.
We assume that CP is conserved on the fundamental level
(all couplings are real) and all flavon vevs are real.
In Appendix \ref{App:Alignment}
we give a superpotential that has the desired flavon vev directions
as a solution and also fixes the phases of the vevs up to a few
discrete choices. For the doublets we find the vev alignments
%%%%%%%%%%%%%%%%%%%%%%%%%%%%
\begin{equation}
\label{eq:2FlavonAlignment}
\begin{split}
\langle \psi^{\prime} \rangle &= \begin{pmatrix} 1 \\ 0 \end{pmatrix} \psi^{\prime}_{0} \;, \quad
\langle \psi^{\prime \prime} \rangle = \begin{pmatrix} 0 \\ 1 \end{pmatrix} \psi^{\prime \prime}_{0} \;, \quad\\
\langle \ti\psi^{\prime} \rangle &= \begin{pmatrix} 1 \\ 0 \end{pmatrix} \ti\psi^{\prime}_{0} \;, \quad
\langle \ti\psi^{\prime \prime} \rangle = \begin{pmatrix} 0 \\ 1 \end{pmatrix} \ti\psi^{\prime \prime}_{0}
\end{split}
\end{equation}
%%%%%%%%%%%%%%%%%%%%%%%%%%%%%%%%%
%
Furthermore we have introduced six flavons
in one-dimensional representations of $T^{\prime}$
which receive all non-vanishing (and real) vevs
%%%%%%%%%%%%%%%%%%%%%%%%%%%%%%%%%%%%
\begin{align}
\label{eq:1FlavonAlignment}
\langle \zeta^{\prime} \rangle &= \zeta^{\prime}_{0} \;, \quad
\langle \zeta^{\prime \prime} \rangle = \zeta^{\prime \prime}_{0} \;, \quad
\langle \ti\zeta^{\prime}\rangle = \ti\zeta^{\prime}_{0} \;, \quad
\langle \ti\zeta^{\prime \prime} \rangle = \ti\zeta^{\prime \prime}_{0} \;, \quad
\langle \rho \rangle = \rho_{0} \;, \quad
\langle \ti\rho \rangle = \ti\rho_{0} \;.
\end{align}
%%%%%%%%%%%%%%%%%%%%%%%%%%
%
All flavons including their quantum numbers are
summarised in Tab.\ \ref{tab:FlavonMatter}.
As we will see soon the flavon field
$\zeta^{\prime}$ does not directly couple to the
matter sector. Nevertheless, we mention it here
because it behaves differently than the auxiliary
$\epsilon$ flavons which we have introduced to get
the desired alignment and make all vevs real, see
Appendix \ref{App:Alignment}.
%%%%%%%%%%%%%%%%%%%%%%%%%%%%%
\begin{table}
\centering
\begin{tabular}{cc ccccc cccccc c}
\toprule
 & $\ti \phi $& $\ti\psi^{\prime\prime}$& $\ti\psi^{\prime } $& $\ti\zeta^{\prime\prime} $& $\ti\zeta^{\prime } $& $\phi$ & $\psi^{\prime\prime} $& $\psi^{\prime } $& $\zeta ^{\prime\prime}$& $\zeta ^{\prime }$& $\xi$ & $\rho$ & $\ti\rho $\\
 \midrule
 $SU(5)$ & $\mathbf{1}$ & $\mathbf{1}$ & $\mathbf{1}$ & $\mathbf{1}$ & $\mathbf{1}$ & $\mathbf{1}$  & $\mathbf{1}$& $\mathbf{1}$& $\mathbf{1}$ & $\mathbf{1}$ & $\mathbf{1}$ & $\mathbf{1}$ & $\mathbf{1}$ \\
 $T^\prime$ &  $\mathbf{3}$ & $\mathbf{2^{\prime\prime}}$ & $\mathbf{2^{\prime }}$ & $\mathbf{1^{\prime\prime}}$ & $\mathbf{1^{\prime}}$ & $\mathbf{3}$ & $\mathbf{2^{\prime\prime}}$ & $\mathbf{2^{\prime }}$ & $\mathbf{1^{\prime\prime}}$ & $\mathbf{1^{\prime}}$ & $\mathbf{3}$ & $\mathbf{1}$ & $\mathbf{1}$ \\
 $U(1)_R$ & 0 & 0 & 0 & 0 & 0 & 0 & 0 & 0 & 0 & 0 & 0 & 0 &0 \\
 $Z_{12}^u$& 0& 3& 9& 0& 0& 6& 3& 9& 6& 0& 6& 6& 6\\
 $Z_{8}^d$ & 0& 0& 0& 0& 0& 2& 1& 7& 6& 4& 4& 4& 4\\
 $Z_{8}^\nu $  &4& 1& 7& 0& 0& 2& 7& 1& 6& 4& 0& 0& 0\\
 $Z_{8} $ &4& 7& 5& 4& 0& 2& 5& 3& 6& 4& 4& 4& 4\\
 $Z_{6} $ &4& 4& 2& 4& 2& 0& 3& 3& 0& 0& 0& 0& 0\\
 $Z_{6}^{\prime} $ &4& 4& 2& 4& 2& 3& 0& 0& 0& 0& 0& 0& 0\\
 $Z_{4} $ &0& 2& 2& 0& 0& 0& 3& 1& 2& 0& 0& 0& 0\\
\bottomrule
\end{tabular}
\caption{\label{tab:FlavonMatter}
Flavon fields coupling to the matter sector including their quantum numbers.
In fact, $\zeta^{\prime}$ does not couple directly to the matter fields, but
it behaves very similar like the other flavons and not like the auxiliary flavons
$\epsilon$ which will be introduced in Appendix \ref{App:Alignment}.
}
\end{table}
%%%%%%%%%%%%%%%%%%%%%%%%%%%%%%%%

\section{The Quark and Charged Lepton Sector}

 In this section we describe the superpotential of the quark and
charged lepton content of the chiral superfields of the model under
study. We will consider the three generations of matter fields in
the usual $\bar{\mathbf{5}}$ and $\mathbf{10}$, five and ten- dimensional,
representations of $SU(5)$, $\bar{F} = (d^c,L)_L$ and $T =
(q,u^c,e^c)_L$. The elements of the Yukawa coupling matrices are
generated dynamically through a number of effective operators which
structure is tightly related to the matter fields assignment under
the $T^{\prime}$ discrete symmetry. Indeed   the Yukawa coupling
matrices can be written only after the breaking of the $T^{\prime}$
discrete symmetry. As will be clear
soon, in this description CP violation in the quark and charged
lepton sector is entirely due to geometrical origin, specifically
from the use of the spinorial representation of the $T^{\prime}$
group. Finally, in this section we will present a $\chi^2$ fit
analysis that has been performed by us to get the low energy masses
and mixing parameters in the quark and charged lepton sector.
We show as well that the the simple CKM phase sum rule from
\cite{Antusch:2009hq} can be applied here.

\subsection{Effective Operators and Yukawa Matrices}

 Before we come to the effective operators
which will give us the Yukawa couplings we first fix the conventions
used for the Yukawa matrices. Throughout this paper we will use the
RL convention, i.e.,
%%%%%%%%%%%%%%%%%%%%%%%%%%%%
\begin{equation}
 - \mathcal{L} = Y_{ij} \overline{f_R^i}  f_L^j H + \text{H.c.}
\end{equation}
%%%%%%%%%%%%%%%%%%%%%%%%
%
or in other words we have to diagonalise the combination
$Y^\dagger Y$. Keep also in mind that $\bar{F} = (d^c,L)_L$
and $T = (q,u^c,e^c)_L$.

We restrict ourselves to effective operators up to mass
dimension seven.
These operators generate Yukawa couplings
of the order of $10^{-5}$ or smaller
(see our fit results in Tab.\ \ref{Tab:FitResults}).
Higher dimensional operators hence can be expected to
give only negligible corrections.

After integrating out the heavy messenger fields,
see Appendix \ref{App:Messengers}, we obtain
the effective operators
%%%%%%%%%%%%%%%%%%%%%%%%%%%%%
\begin{equation}
\begin{split}
\mathcal{W}_{Y_{u}} & =
    y^{(u)}_{33} H_{5}^{(1)}  T_{3} T_{3}
  + \frac{y^{(u)}_{23}}{\Lambda_u^{2}} (T_{a} \ti \phi)_{2^{\prime}} H_{5}^{(2)} (T_{3} \ti \psi^{\prime \prime})_{2^{\prime \prime}}
  + \frac{y^{(u)}_{22}}{\Lambda_u^{3}} (T_{a} \ti \psi^{\prime \prime})_{3} (H_{5}^{(1)} \ti \zeta^{\prime})_{1^{\prime}} ( T_{a} \ti \psi^{\prime \prime})_{3} \\
& + \frac{y^{(u)}_{21}}{\Lambda_u^{4}} (T_{a} \ti \phi)_{2^{\prime}} (H_{5}^{(1)}
\ti \zeta^{\prime})_{1^{\prime}} (\ti \psi^{\prime} (T_a \ti
\psi^{\prime})_3)_{2^{\prime}}
  + \frac{y^{(u)}_{11}}{\Lambda_u^{4}} ((T_{a} \ti \phi)_{2}  \ti \zeta^{\prime \prime})_{2^{\prime \prime}} H_{5}^{(3)}( \ti \zeta^{\prime \prime} (T_{a} \ti \phi)_{2^{\prime \prime}})_{2^{\prime}} \;,
  \label{eq:Wu}
\end{split}
\end{equation}
%%%%%%%%%%%%%%%%%%%%%%%%%
%
which give the up-type quark Yukawa
matrices after the flavons developed their vevs.
Here $\Lambda_u$ stands for the messenger scale
suppressing the non-renormalisable operators
in the up-sector and in the down-sector
we will introduce $\Lambda_d$ correspondingly.
We have also given the $T^{\prime}$ contractions as
indices on the round brackets. Note that in general
there are many different contractions possible
(for $T^{\prime}$ and to a less degree for $SU(5)$)
which give different results. Nevertheless, we have
specified in Appendix \ref{App:Messengers} the fields
mediating the non-renormalisable operators which
transform in a specific way under $T^{\prime}$
such that we pick up only the contractions which we
want.

Multiplying the $T^{\prime}$ and $SU(5)$ indices out we obtain for
the up-type quark Yukawa matrix at the GUT scale (which is roughly
equal to the scale of $T^{\prime}$ breaking)
%%%%%%%%%%%%%%%%%%%%%%%%%%%
\begin{equation}
Y_u = \begin{pmatrix}
\bar \omega a_u & \ci b_u & 0 \\
\ci b_u &  c_u & \omega d_u \\
0 & \omega d_u  & e_u
\end{pmatrix} \;,
\end{equation}
%%%%%%%%%%%%%%%%%%%%%%%%%
%
where $\omega = (1+\ci)/\sqrt{2}$ and $\bar{\omega} =
(1-\ci)/\sqrt{2}$. The parameters $a_u$, $b_u$, $c_u$, $d_u$ and
$e_u$ are (real) functions of the underlying parameters. Note at
this point, that the phases of the flavon vevs have to be fixed.
Otherwise the coefficients in the Yukawa matrix are complex
parameters and we would not be able to make definite predictions
anymore.

For the down-type quarks and charged leptons (remember that those
two sectors are closely related in $SU(5)$) we find for the
superpotential
%%%%%%%%%%%%%%%%%%%%%%%%%%%%%%%%%%
\begin{equation}
\begin{split}
\mathcal{W}_{Y_{d,\ell}} &=
    \frac{y^{(d)}_{33}}{\Lambda_d^{2}} ((\bar H_{5}^{(2)}\bar{F} )_3  \phi)_{1^{\prime}} (H_{24}^{\prime \prime} T_{3})_{1^{\prime \prime}}
  + \frac{y^{(d)}_{22}}{\Lambda_d^{3}} ((\phi T_a)_{2^{\prime}} H_{24}^{\prime \prime})_2 (\psi^{\prime}(\bar H_{5}^{(1)} \bar{F} )_{3}  )_2 \\
& + \frac{y^{(d)}_{12}}{\Lambda_d^{4}} (((T_a \ti H_{24}^{\prime \prime})_{2^{\prime \prime}} ( \bar F \psi^{\prime})_{2^{\prime
\prime}} )_3 \psi^{\prime} )_{2^{\prime\prime}}  ( \bar{H}_{5}^{(3)} \psi^{\prime})_{2^{\prime}}
  + \frac{y^{(d)}_{21}}{\Lambda_d^{4}} ((\bar F \psi^{\prime})_{2^{\prime \prime}} (\zeta^{\prime \prime}  \bar{H}_{5}^{(1)})_{1^{\prime \prime} } \zeta^{\prime \prime})_2  (T_a \phi)_{2} \\
&  + \frac{y^{(d)}_{11}}{\Lambda_d^{4}} ( (\bar{F} \psi^{\prime \prime})_{2^{\prime}} (H_{24}^{\prime \prime}\psi^{\prime \prime})_{2^{\prime}} \bar H_{5}^{\prime \prime} )_{1^{\prime}} (T_a \psi^{\prime \prime})_{1^{\prime \prime}} \;,
\label{eq:Ydown}
\end{split}
\end{equation}
%%%%%%%%%%%%%%%%%%%%%%%%%
%
where we have again specified the $T^{\prime}$ contractions. From
this superpotential and considering the correct $SU(5)$
contractions, which we could not display here for the sake of
readability, we get the down-type quark and charged lepton Yukawa
matrices
%%%%%%%%%%%%%%%%%%%%%%%%%%%%%%%%
\begin{equation}
Y_d = \begin{pmatrix}
  \omega \, a_d\, & \ci b_d^{\prime} & 0 \\
  \bar{\omega} \, b_d & c_d & 0 \\
 0 &  0 & d_d
\end{pmatrix} \;\;  \text{and} \quad Y_e = \begin{pmatrix}
 -\frac{3}{2} \, \omega \, a_d & \bar{\omega} \, b_d & 0 \\
 6 \, \ci \, b_d^{\prime} &  6\, c_d & 0 \\
 0 &  0 & -\frac{3}{2}\, d_d
\end{pmatrix} \;, \label{eq:Yd}
\end{equation}
%%%%%%%%%%%%%%%%%%%%%%%%%%%
%
where $a_d$, $b_d$, $b_d^{\prime}$, $c_d$ and
$d_d$ are (real) functions of the underlying
parameters.

Note that the prediction from the minimal
$SU(5)$ model $Y_d = Y_e^T$ is broken.
Indeed it has to be broken to get realistic
fermion masses. For the second generation
this is known for a long time \cite{Georgi:1979df}.
In some recent work \cite{Antusch:2009gu}
some new relations to fix this issue were
proposed. From those we will use here
$y_\tau/y_b = -3/2$ and $y_\mu/y_s \approx 6$
where $y_\tau$, $y_\mu$, $y_b$ and $y_s$
stand for the eigenvalues of the Yukawa matrices
associated to the masses of the $\tau$, the $\mu$,
the $b$ and the $s$ quark respectively.
Furthermore it was shown in \cite{Marzocca:2011dh}
(see also \cite{Antusch:2011qg}) that those
new $SU(5)$ Clebsch--Gordan coefficients
might also give a large reactor neutrino mixing
angle $\theta_{13}$. For the current paper
we have chosen one of the possible combinations
given in \cite{Marzocca:2011dh} but we remark
that in principle also other combinations are still
possible which might be realised in another
unified flavour model  with a similar good fit
to the fermion masses and mixing angles.

\subsection{Fit Results and the CKM Phase Sum Rule}

In the last section we have discussed the structure of the Yukawa
matrices in the quark and the charged lepton sector. These matrices
have five free parameters, which in principle can be fitted to the
low-energy mass and mixing parameters using the renormalisation
group. But doing so one has to take into account SUSY threshold
corrections \cite{SUSYthresholds} which modify the masses and mixing
angles significantly. For example without including them, the GUT
scale Yukawa coupling ratio, $y_\tau/y_b$, would be roughly 1.3
which is not close to the usual GUT prediction of~1.
There is a large amount of literature on how to use
SUSY threshold corrections to get $b-\tau$ Yukawa
unification, for recent papers see, for instance,
\cite{Antusch:2008tf, Antusch:2009gu, Profumo:2003ema}.
From these studies it is known that in 
order to get $b-\tau$ Yukawa unification, it is
necessary to either consider a negative $\mu$-term or to have 
a very high - $\mathcal{O}$(10 TeV), SUSY scale.
Nevertheless, we will not use unifcation but instead we use
the recently proposed GUT scale relation $y_\tau/y_b = 3/2$
induced by the vev of an adjoint of $SU(5)$
\cite{Antusch:2009gu}, which is viable in a large 
region of the parameter space even 
in constrained MSSM scenarios.

Due to the importance of the threshold corrections
for our fit we briefly revise the most important formulas
which also defines our parameterisation.
 In \cite{Antusch:2011sq}
the approximate matching conditions at the SUSY scale,
$M_{\text{SUSY}}$,
%%%%%%%%%%%%%%%%%%%%%%%%%%
\begin{align}
   y^{\text{SM}}_{e,\mu,\tau} &= (1 + \epsilon_l \tan\beta)~ y^{\text{MSSM}}_{e,\mu,\tau} \cos \beta \;,\\
   y^{\text{SM}}_{d,s} &= (1 + \epsilon_q \tan\beta)~ y^{\text{MSSM}}_{d,s} \cos \beta \;,\\
   y^{\text{SM}}_{b} &= (1 + (\epsilon_q + \epsilon_A) \tan\beta)~ y^{\text{MSSM}}_b \cos \beta \;,
\end{align}
%%%%%%%%%%%%%%%%%%%%%%%
%
for the Yukawa couplings and
%%%%%%%%%%%%%%%%%%%%%%
\begin{align}
   \theta^{\text{SM}}_{i3} &= \frac{1 + \epsilon_q \tan\beta}{1 + (\epsilon_q + \epsilon_A) \tan\beta}~ \theta^{\text{MSSM}}_{i3} \;, \\
   \theta^{\text{SM}}_{12} &= \theta^{\text{MSSM}}_{12} \;, \\
   \delta^{\text{SM}}_{\text{CKM}} &= \delta^{\text{MSSM}}_{\text{CKM}} \;,
\end{align}

%%%%%%%%%%%%%%%%%%%%%
%
for the quark mixing parameters were given,
where the SUSY threshold corrections are parameterised
in terms of the three parameters $\epsilon_l$,
$\epsilon_q$ and $\epsilon_A$. We will adopt
this parameterisation neglecting $\epsilon_l$,
which is usually one order smaller than
$\epsilon_q$ \cite{Antusch:2008tf}.
Furthermore we want to assume, that SUSY
is broken similar to the constrained MSSM
scenario with a positive $\mu$ parameter
and hence
we adopt the recently proposed GUT relation
$y_\tau/y_b = 3/2$ for the third generation,
as mentioned earlier. For the second
generation we use $y_\mu/y_s \approx 6$
\cite{Antusch:2009gu}.

We have fixed the SUSY scale to 750~GeV,
the GUT scale to $2 \times 10^{16}$~GeV and
$\tan \beta$ to 35. Therefore we have to
fit the ten parameters in the Yukawa matrices
and the two parameters from the SUSY threshold
corrections to the thirteen low energy observables
in the quark and the charged lepton sector (nine
masses, three mixing angles and one phase),
so that we have one prediction (degree of freedom).
%%%%%%%%%%%%%%%%%%%%%%%%%%%%%%%
\begin{table}
\begin{center}
\begin{tabular}{cc}
\toprule
Parameter & Value \\ \midrule
$a_u$ & $ 5.81 \cdot 10^{-6}$ \\
$b_u$ & $-9.96 \cdot 10^{-5}$ \\
$c_u$ & $-8.55 \cdot 10^{-4}$ \\
$d_u$ & $1.99 \cdot 10^{-2}$\\
$e_u$ & $0.525$ \\
\midrule
$a_d$ & $-2.82 \cdot 10^{-5}$ \\
$b_d$ & $-5.73 \cdot 10^{-4}$\\
$b'_d$ & $-5.09 \cdot 10^{-4}$ \\
$c_d$ & $2.50 \cdot 10^{-3}$ \\
$d_d$ & $1.82 \cdot 10^{-1}$ \\
\midrule
$\epsilon_q \tan \beta$ & $ 0.1788$ \\
$\epsilon_A \tan \beta$ & $-0.0001$ \\
\bottomrule
\end{tabular}
\caption{Values of the effective parameters of the quark and
charged lepton Yukawa matrices for $\tan \beta = 35$ and
$M_{\text{SUSY}} = 750$~GeV. The two parameters $\epsilon_q$ and
$\epsilon_A$ parameterise the SUSY threshold corrections. The
numerical values are determined from a $\chi^2$-fit to experimental
data with a lowest $\chi^2$ per degree of freedom of 2.76.
\label{Tab:Parameters}}
\end{center}
\end{table}
%%%%%%%%%%%%%%%%%%%%%%%%%
%%%%%%%%%%%%%%%%%%%%%%%%
\begin{table}
\begin{center}
\begin{tabular}{cccc}
\toprule
Quantity (at $m_t(m_t)$)& Experiment & Model & Deviation \\ \midrule
$y_\tau$ in $10^{-2}$   & 1.00          & $0.99$ & -0.388 \\
$y_\mu$ in $10^{-4}$    & 5.89          & $5.90$ &  0.044 \\
$y_e$ in $10^{-6}$  & 2.79          & $2.79$ & -0.003 \\
\midrule
$y_b$ in $10^{-2}$      & $1.58 \pm 0.05$   & $1.57$ & -0.157 \\
$y_s$ in $10^{-4}$  & $2.99 \pm 0.86$   & $2.57$ & -0.484 \\
$y_s/y_d$           & $18.9 \pm 0.8$    & $18.9$ & -0.012 \\
\midrule
$y_t$               & $0.936 \pm 0.016$ & $0.936$&  0.0001  \\
$y_c$ in $10^{-3}$      & $3.39 \pm 0.46$   & $2.79$ & -1.317  \\
$y_u$ in $10^{-6}$      & $7.01^{+2.76}_{-2.30}$& $7.01$ & -0.0003,  \\
\midrule
$\theta_{12}^{\text{CKM}}$ & $0.2257^{+0.0009}_{-0.0010}$ & $0.2257$ & -0.0107 \\[0.3pc]
$\theta_{23}^{\text{CKM}}$ & $0.0415^{+0.0011}_{-0.0012}$ & $0.0416$ & 0.1268  \\[0.1pc]
$\theta_{13}^{\text{CKM}}$ & $0.0036 \pm 0.0002$          & $0.0036$ & 0.2043  \\[0.1pc]
$\delta_{\text{CKM}}$ & $1.2023^{+0.0786}_{-0.0431}$      & $1.2610$ & 0.7465  \\
\bottomrule
\end{tabular}
\caption{Fit results for the quark Yukawa couplings and mixing and
the charged lepton Yukawa couplings at low energy compared to
experimental data. The values for the Yukawa couplings are extracted
from \cite{Xing:2007fb}, the ratio $y_s/y_d$ is taken from
\cite{Leutwyler:2000hx} and the CKM parameters from \cite{PDG10}.
Note that the experimental uncertainty on the charged lepton Yukawa
couplings are negligible small and we have assumed a relative
uncertainty of 3~\% for them. The $\chi^2$ per degree of freedom
is 2.76. A pictorial representation of the agreement between our
fit and experiment can be found as well in
Fig.~\ref{Fig:FitResultsPlot}. \label{Tab:FitResults}}
\end{center}
\end{table}
%%%%%%%%%%%%%%%%%%%%%%%%%%%
%%%%%%%%%%%%%%%%%%%%%%%%%%
\begin{figure}
\centering
\includegraphics[scale=1]{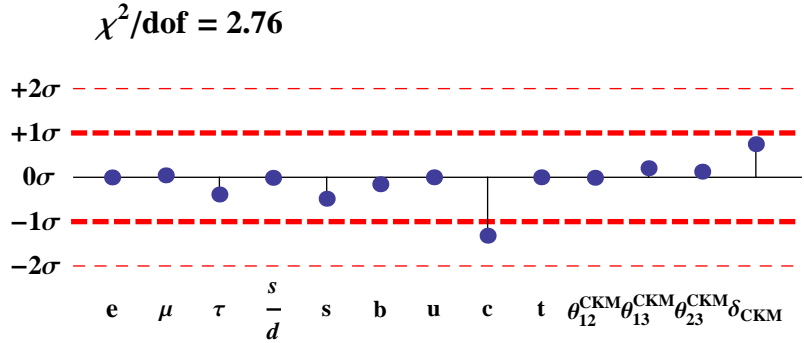}
\caption{Pictorial representation of the deviation
of our fit from low energy experimental data
for the charged lepton Yukawa couplings and quark Yukawa
couplings and mixing parameters.  The deviations of the
charged lepton masses are given in~3\% while all other
deviations are given in units of standard deviations $\sigma$.
\label{Fig:FitResultsPlot} }
\end{figure}
%%%%%%%%%%%%%%%%%%%%%%%%%%%%%%
%

The RGE running and diagonalisation of the matrices
was done using the {\tt REAP}
package \cite{Antusch:2005gp}. Performing a $\chi^2$
fit we have found as minimum the results listed in Tab.\
\ref{Tab:Parameters} for the parameters and in Tab.\
\ref{Tab:FitResults} and in Fig.~\ref{Fig:FitResultsPlot}
we have presented the results of the fit for the low energy
observables compared to the experimental results.
Note that we have assumed an uncertainty of 3\%
on the Yukawa couplings for the charged leptons.
Their experimental uncertainty is much smaller, so that
their theoretical uncertainty (Accuracy of RGEs, neglecting
SUSY threshold corrections for the leptons, NLO effects, ...)
is much bigger, which we estimate to be 3\%.

We find good agreement between our model and
experimental data with a minimal $\chi^2$ per degree
of freedom of 2.76. In fact this agreement is not
accidental. We have chosen the $SU(5)$ coefficients
such that, we expect good agreement and we have also
enough free parameters to fix the mixing angles.
In other words one could determine the eigenvalues and
mixing angles from the data and then the CKM phase would
be a prediction. But as we will demonstrate now, the choice
for our phases in the Yukawa matrices was done in such a
way, that we can expect a good prediction for the CKM
phase as well.

We will show in the following that the sum rule given in
\cite{Antusch:2009hq} can be used here.
To apply the sum rule we have to find approximate
expressions for the complex mixing angles (see
\cite{Antusch:2009hq}).
For the rest of the subsection we will use the
the notation of \cite{Antusch:2009hq} which we just
briefly summarise here for convenience.
The CKM matrix $U_{\text{CKM}}$ can be written as
\begin{equation}
U_{\text{CKM}} =  U_{u_L} U_{d_L}^\dagger = (U^{u_L}_{23} U^{u_L}_{13} U^{u_L}_{12})^\dagger U^{d_L}_{23} U^{d_L}_{13} U^{d_L}_{12} \;,
\end{equation}
where the matrices $U_{u_L}$ and $U_{d_L}$ diagonalise
the up- and down-type quark mass matrices and the unitary
matrix
\begin{equation}
U_{12}= \begin{pmatrix}
  \cos \theta_{12} & \sin \theta_{12} \, \text{e}^{-\ci \delta_{12}} & 0\\
  - \sin \theta_{12} \, \text{e}^{\ci \delta_{12}}  &  \cos \theta_{12} & 0\\
  0&0&
  \end{pmatrix}\,.
\end{equation}
The matrices $U_{13}$ and $U_{23}$ are given by analogous expressions.

We find at leading order for the respective
mixing angles and phases
%%%%%%%%%%%%%%%%%%%%%%%%%%
\begin{align}
 \theta_{12}^d \; \text{e}^{-\ci \delta_{12}^{d}} &= \left| \frac{b_d}{c_d} \right| \text{e}^{- \ci \frac{7 \pi}{4}} \;, \;\; \theta_{13}^d = \theta_{23}^d = 0 \;, \label{eq:ApproxDown}\\
 \theta_{12}^u \; \text{e}^{-\ci \delta_{12}^{u}} &\approx \left| \frac{b_u}{\sqrt{2} c_u} \right| \text{e}^{- \ci \frac{5 \pi}{4} } \;, \;\;
\theta_{23}^u \; \text{e}^{-\ci \delta_{23}^{u}} = \left|
\frac{d_u}{e_u} \right| \text{e}^{- \ci \frac{5 \pi}{4} } \;, \;\;
\theta_{13}^u \; \text{e}^{-i\delta_{13}^{u}} = \left| \frac{b_u
d_u}{e_u^2} \right| \text{e}^{- \ci \frac{\pi}{4}} \label{eq:ApproxUp} \;,
\end{align}
%%%%%%%%%%%%%%%%%%%%%%%%
%
where we have used for $\theta_{12}^u$ that
$d_u^2 \approx -1/2 c_u$ and $e_u \approx 0.5$ from
our fit. So we see that $\delta_{12}^u$ is
not simply $\pi/2$ as one would expect from a quick first
inspection. Note also that the phase sum rule
was derived for $\theta_{13}^u = \theta_{13}^d = 0$,
which is not exactly true in our case for $\theta_{13}^u$.
But in fact it is sufficient, that
$\theta_{13}^u \ll \theta_{12}^u \theta_{23}$
which is fulfilled here.

The angle $\alpha$ in the CKM unitarity triangle is experimentally
measured to be $\alpha =
(90.7^{+4.5}_{-2.9} )^\circ$ \cite{CKM}  for which
the sum rule
%%%%%%%%%%%%%%%%%%%%%%%%%%%%%
\begin{equation}
 \alpha \approx \delta_{12}^d - \delta_{12}^u\;,
\end{equation}
%%%%%%%%%%%%%%%%%%%%%%%%%
%
was given in \cite{Antusch:2009hq}. Plugging in our
approximate analytical expressions for
$\delta_{12}^{d/u}$, eqs.\ \eqref{eq:ApproxDown}
and \eqref{eq:ApproxUp},
we find that $\alpha \approx \pi/2$ and our model is
in good agreement with experimental data as we have
also seen it before from our numerical fit.

\section{Neutrino Sector}

The model includes three heavy right-handed Majorana
neutrino fields $N$ which are singlets under $SU(5)$
and a triplet under $T^{\prime}$. Through the type I
seesaw mechanism \cite{seesaw} we generate light
neutrino masses. The neutrino sector is described by
the following terms in the superpotential
%%%%%%%%%%%%%%%%%%%%%%%%%%%
\begin{equation}
\mathcal{W}_\nu = \lambda_1 N N \xi + N N (\lambda_2 \rho + \lambda_3 \tilde{\rho})
 + \frac{y_{\nu}}{\Lambda} (N \bar F )_1 (H_5^{(2)} \rho)_1
 + \frac{\tilde{y}_{\nu}}{\Lambda}  (N \bar F)_1 (H_5^{(2)} \tilde{\rho})_1
 \;, \label{eq:Wnu}
\end{equation}
%%%%%%%%%%%%%%%%%%%%%%%%%
%
where we have given the $T^{\prime}$ contractions
as indices at the brackets for non-renormalisable terms
and from now on $\Lambda$ labels a generic
messenger scale.
Note that the contraction of three triplets in general is not
unique, see also Tab.\ \ref{tab:CGs}, because the product
of two triplets contains a symmetric and an antisymmetric
triplet. But since we multiply here two $N$ with each other
only the symmetric combination gives a non-vanishing
contribution. In the following we will discuss the
phenomenological implications of this superpotential
(including corrections from the charged lepton sector).

\subsection{The Neutrino Mass Spectrum}

From eq.\ \eqref{eq:Wnu} we obtain for the mass matrix
for the right-handed neutrinos and the Dirac neutrino
mass matrix
%%%%%%%%%%%%%%%%%%%%%%%%%%%%%%%%%%%
\begin{equation}
 M_R = \begin{pmatrix} 2 Z+X & -Z & -Z \\ -Z & 2 Z & -Z+X \\ -Z & - Z+X  & 2 Z \end{pmatrix} \;,\quad
 M_D = \begin{pmatrix} 1 & 0 & 0 \\ 0 & 0 & 1 \\ 0 & 1  & 0 \end{pmatrix}
\frac{\rho^{\prime}}{\Lambda}\; ,
\end{equation}
%%%%%%%%%%%%%%%%%%%%%%%%%%%%%%%%%%%%%
%
where $X$, $Z$ and $\rho^{\prime}$ are real
parameters depending on the couplings and the vevs in eq.\ \eqref{eq:Wnu}.
The right-handed neutrino mass matrix $M_R$ is diagonalised
by the tri-bimaximal mixing (TBM) matrix \cite{tri1}
%%%%%%%%%%%%%%%%%%%%%%%%%%%%%%%%
\begin{equation}
\TBM =
\begin{pmatrix}
\sqrt{2/3} & \sqrt{1/3} & 0 \\
-\sqrt{1/6} & \sqrt{1/3}  &  -\sqrt{1/2}\\
-\sqrt{1/6} &  \sqrt{1/3}  & \sqrt{1/2}
\end{pmatrix} \;,
\end{equation}
%%%%%%%%%%%%%%%%%%%%%%%
such that the heavy RH neutrino masses read:
%%%%%%%%%%%%%%%%%%%%%%%%
\begin{equation}
\label{eq:MR} \TBM^T\, M_R\,\TBM = D_{N} = \diag(3Z + X,X,3Z - X) =
\diag(M_1 e^{\ci \phi_1},M_2 e^{\ci \phi_2},M_3 e^{\ci \phi_3})\;,~M_{1,2,3} > 0\,,
\end{equation}
%%%%%%%%%%%%%%%%%%%%%%
%
where
\begin{eqnarray}
M_1 &=& |X+3Z| \;\equiv\;|X|\,|1+\alpha e^{\ci \phi}|,\,\,\,\phi_1={\rm arg}(X+3Z)
\label{M1}\\
M_2 &=& |X|,\,\,\,\,\,\phi_2={\rm arg}(X)
\label{M2}\\
M_3 &=& |X-3Z| \;\equiv\;|X|\,|1-\alpha e^{\ci \phi}|,\,\,\,\,\,
\phi_3={\rm arg}(3Z-X)\,.
\label{M3}
\end{eqnarray}
%%%%%%%%%%%%%%%%%%%%%%
%
Here $\alpha\equiv|3Z/X| > 0$ and $\phi\equiv{\rm arg}(Z)-{\rm
arg}(X)$.
Since $X$ and $Z$ are real parameters, the phases
$\phi_1$, $\phi_2$, $\phi_3$ and $\phi$ take values 0 or $\pi$.
A light neutrino Majorana mass term is generated
after electroweak symmetry breaking
via the type I see-saw mechanism:
%%%%%%%%%%%%%%%%%%%%%%%%%%%%%%%%%
\begin{equation}\label{mnuLO}
M_\nu  \;=\;- M_D^T\, M_R^{-1}\,M_D =
U^*_{\nu}\diag\left(m_1,m_2,m_3\right)U^\dagger_{\nu}\,,
\end{equation}
%%%%%%%%%%%%%%%%%%%%%
where
%%%%%%%%%%%%%%%%%%%
\begin{equation}
U_{\nu} = \ci \, \TBM \,\diag\left(e^{\ci \phi_1/2},e^{\ci \phi_2/2},
e^{\ci \phi_3/2}\right) \equiv \ci\, \TBM \,\tilde{Q}\,,~~
\tilde{Q}\equiv \diag\left(e^{\ci \phi_1/2},e^{\ci \phi_2/2},e^{\ci \phi_3/2}\right)\,,
\label{Unu}
%\label{U_matrix}
\end{equation}
%%%%%%%%%%%%%%%%%%%%%%%%%%
%
and $m_{1,2,3} > 0$ are the light neutrino masses,
%%%%%%%%%%%%%%%%%%%%%%%%%%%%%
\begin{equation}
 m_i = \left(\frac{\rho^{\prime}}{\Lambda}\right )^2\,
\frac{1}{M_i}\,,\,\,\,i=1,2,3\,.
\label{numasses}
\end{equation}
%%%%%%%%%%%%%%%%%%%%%%%%%%
%
The phase factor $\ci$ in eq.\ (\ref{Unu})
corresponds to an unphysical
phase and we will drop it in what follows.
Note also that one of the phases $\phi_k$, say $\phi_1$,
is physically irrelevant since it can be considered
as a common phase of the neutrino mixing matrix.
In the following we always set $\phi_1=0$.
This corresponds to the choice $(X + 3Z) > 0$.

 The type of the neutrino mass spectrum in the model
is determined
\footnote{We are following in this part
the similar analysis performed in \cite{CHEMSP09}.}
by the value of the phase $\phi$. Indeed, as it is not
difficult to show, we have:
%%%%%%%%%%%%%%%%%%%%%%%%%%%%%
\begin{equation}
\Delta m^2_{31} \equiv \Delta m^2_{A} =
\frac{1}{|X|^2}\,\left(\frac{\rho^{\prime}}{\Lambda}\right )^4\,
\frac{4\alpha\,\cos\phi}
{\left |1 + \alpha\,e^{\ci \phi}\right |^2
\left |1 - \alpha\,e^{\ci \phi}\right |^2}\,.
\label{Dm231}
\end{equation}
%%%%%%%%%%%%%%%%%%%%%%%%%
%
Thus, for $\cos\phi = +1$, we get $\Delta m^2_{31} > 0$, i.e.,
a neutrino mass spectrum with normal ordering (NO),
while for $\cos\phi = -1$ one has  $\Delta m^2_{31} < 0$, i.e.,
neutrino mass spectrum with inverted ordering (IO).
We have also:
%%%%%%%%%%%%%%%%%%%%%%%%%%%%%
\begin{equation}
\Delta m^2_{21} \equiv \Delta m^2_{\odot} =
\frac{1}{|X|^2}\,\left(\frac{\rho^{\prime}}{\Lambda}\right )^4\,
\frac{\alpha\,(\alpha + 2\cos\phi)}
{\left |1 + \alpha\,e^{\ci \phi}\right |^2}\,.
\label{Dm221}
\end{equation}
%%%%%%%%%%%%%%%%%%%%%%%%%
%
For a given type of neutrino mass spectrum, i.e., for
a fixed $\phi = 0~{\rm or}~\pi$,
a constraint on the parameter $\alpha$ can be obtained
from the requirement that $\Delta m^2_{21} > 0$
and from the data on the ratio:
%%%%%%%%%%%%%%%%%%%%%%%
\begin{equation}
 r = \frac{\Delta m^2_{\odot}}{|\Delta m^2_{A}|} =
\frac{1}{4}\,\left(\alpha + 2\cos\phi \right)
\left(1 -2\alpha\cos\phi + \alpha^2 \right) =
0.032 \pm 0.006 \;.
\label{eq:ratior}
\end{equation}
%%%%%%%%%%%%%%%%%%%%%%%
%
Using the values of $\alpha$ thus found and
the value of, e.g., $\Delta m^2_{21}$,
one can get (for a given type of the spectrum)
the value of the factor in eq.\ (\ref{Dm221}),
$|X|^{-2}(\rho^{\prime}/\Lambda)^4$.
Knowing this factor and  $\alpha$, one can obtain the value of
the lightest neutrino mass, which together with the data
on  $\Delta m^2_{21}$ and  $\Delta m^2_{31(32)}$
allows to obtain the values of the other
two light neutrino masses. Knowing the latter
one can find also the two ratios of the heavy
Majorana neutrino masses.

  In the case of NO neutrino mass spectrum ($\phi = 0$),
there are two values of $\alpha$ which satisfy
equation (\ref{eq:ratior}) for $r = 0.032$:
$\alpha \cong 1.20$ (solution A), and
$\alpha \cong 0.79$ (solution B).
In the case of solution A, as it is not
difficult to show, the phases
%%%%%%%%%%%%%%%%%%%%%%%%%
\begin{equation}
 \phi_2 = 0\,,\quad \phi_3 = 0\,,~~{\rm solution~A~(NO)}\,,
\label{eq:var2var3A}
\end{equation}
%%%%%%%%%%%%%%%%%%%%%%%%%%
%
and the three neutrino masses have the values:
%%%%%%%%%%%%%%%%%%%%%%%%%
\begin{equation}
m_1 \cong 4.44\times 10^{-3}~{\rm eV}\,, m_2 \cong 9.77\times 10^{-3}~{\rm eV}\,,
m_3 \cong 4.89\times 10^{-2}~{\rm eV}\,,~{\rm solution~A~(NO)}\,.
\label{eq:m123A}
\end{equation}
%%%%%%%%%%%%%%%%%%%%%%%%%%
%
Evidently, the spectrum is mildly hierarchical.
The ratios of the heavy Majorana neutrino
masses read: $M_1/M_3 \cong 11.0$ and
$M_2/M_3 \cong 5.0$. Thus, we have
$M_3 < M_2 < M_1$.

 For solution B we find
%%%%%%%%%%%%%%%%%%%%%%%%%
\begin{equation}
 \phi_2 = 0\,,\quad \phi_3 = \pi\,,~~{\rm solution~B~(NO)}\,,
\label{eq:var2var3B}
\end{equation}
%%%%%%%%%%%%%%%%%%%%%%%%%%
%
while for the values of the three neutrino
masses we get:
%%%%%%%%%%%%%%%%%%%%%%%%%
\begin{equation}
m_1 \cong 5.89\times 10^{-3}~{\rm eV}\,, m_2 \cong 1.05\times 10^{-2}~{\rm eV}\,,
m_3 \cong 4.90\times 10^{-2}~{\rm eV}\,,~{\rm solution~B~(NO)}\,.
\label{eq:m123B}
\end{equation}
%%%%%%%%%%%%%%%%%%%%%%%%%%
%
The heavy Majorana neutrino mass ratios
are given by: $M_1/M_3 \cong 8.33$ and
$M_2/M_3 \cong 4.67$. Therefore also in this case
we have $M_3 < M_2 < M_1$.

 For the IO spectrum ($\phi = \pi$),
we find only one value of  $\alpha$ which
satisfies eq.\ (\ref{eq:ratior}) with $r=0.032$:
$\alpha \cong 2.014$. The phases $\phi_2$ and $\phi_3$
take the values: $\phi_2 = \pi$, $\phi_3 = 0$.
The light neutrino masses read:
%%%%%%%%%%%%%%%%%%%%%%%%%
\begin{equation}
m_1 \cong 5.17\times 10^{-2}~{\rm eV}\,,
m_2 \cong 5.24\times 10^{-2}~{\rm eV}\,,
m_3 \cong 1.74\times 10^{-2}~{\rm eV}\,,~~{\rm (IO)}\,,
\label{eq:m123IO}
\end{equation}
%%%%%%%%%%%%%%%%%%%%%%%%%%
%
i.e., the light neutrino mass spectrum
is not hierarchical exhibiting only partial
hierarchy.
For the heavy Majorana neutrino mass ratios
we obtain: $M_1/M_2 \cong 1.014$ and
$M_3/M_2 \cong 3.01$. Thus, in this case
$N_1$ and $N_2$ are quasi-degenerate in mass:
$M_1 \cong M_2 < M_3$.

%%%%%%%%%%%%%%%%%%%%%%%%%%%%
\begin{figure}
   \begin{center}
\subfigure
   {\includegraphics[width=7cm]{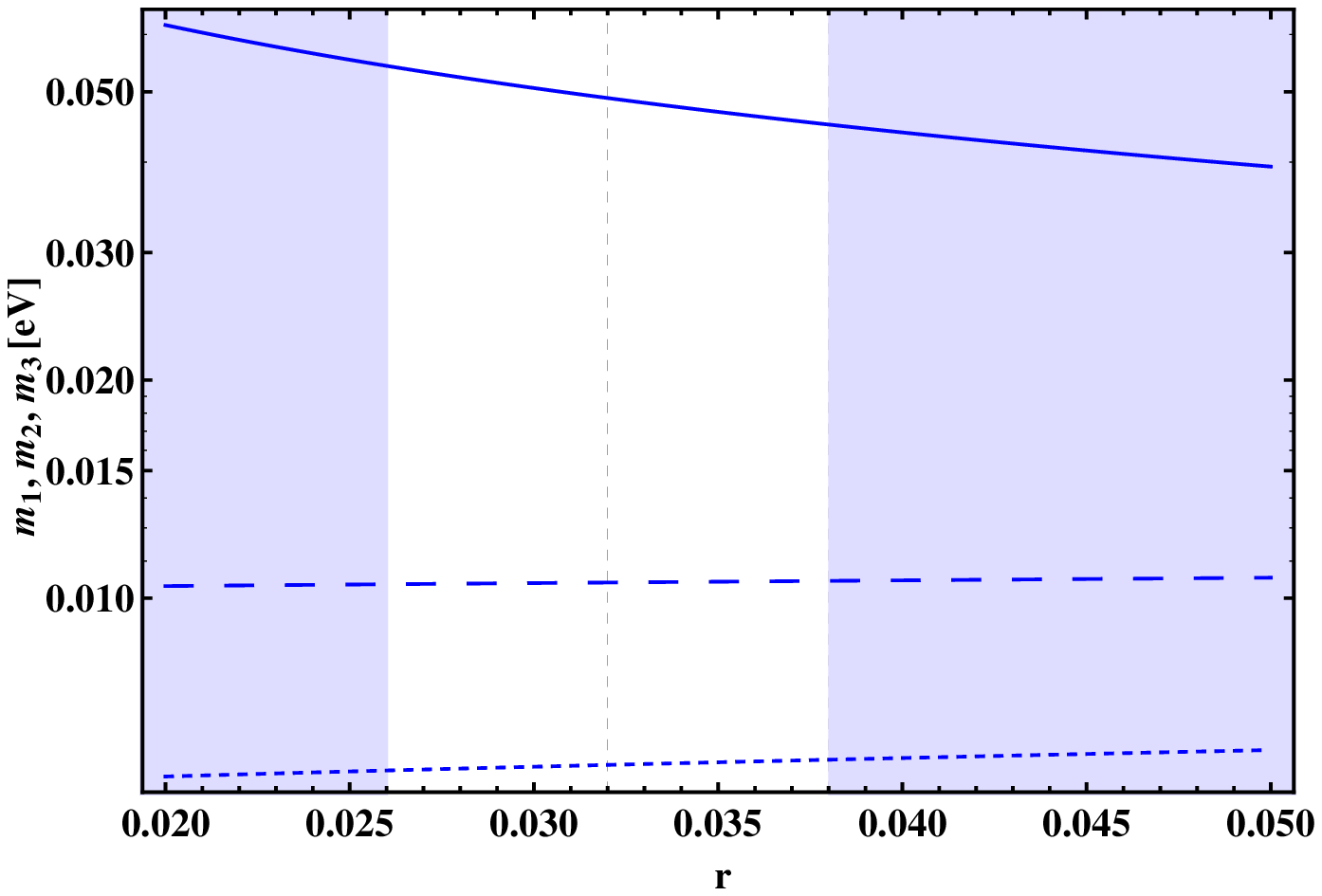}}
 \vspace{5mm}
 \subfigure
   {\includegraphics[width=7cm]{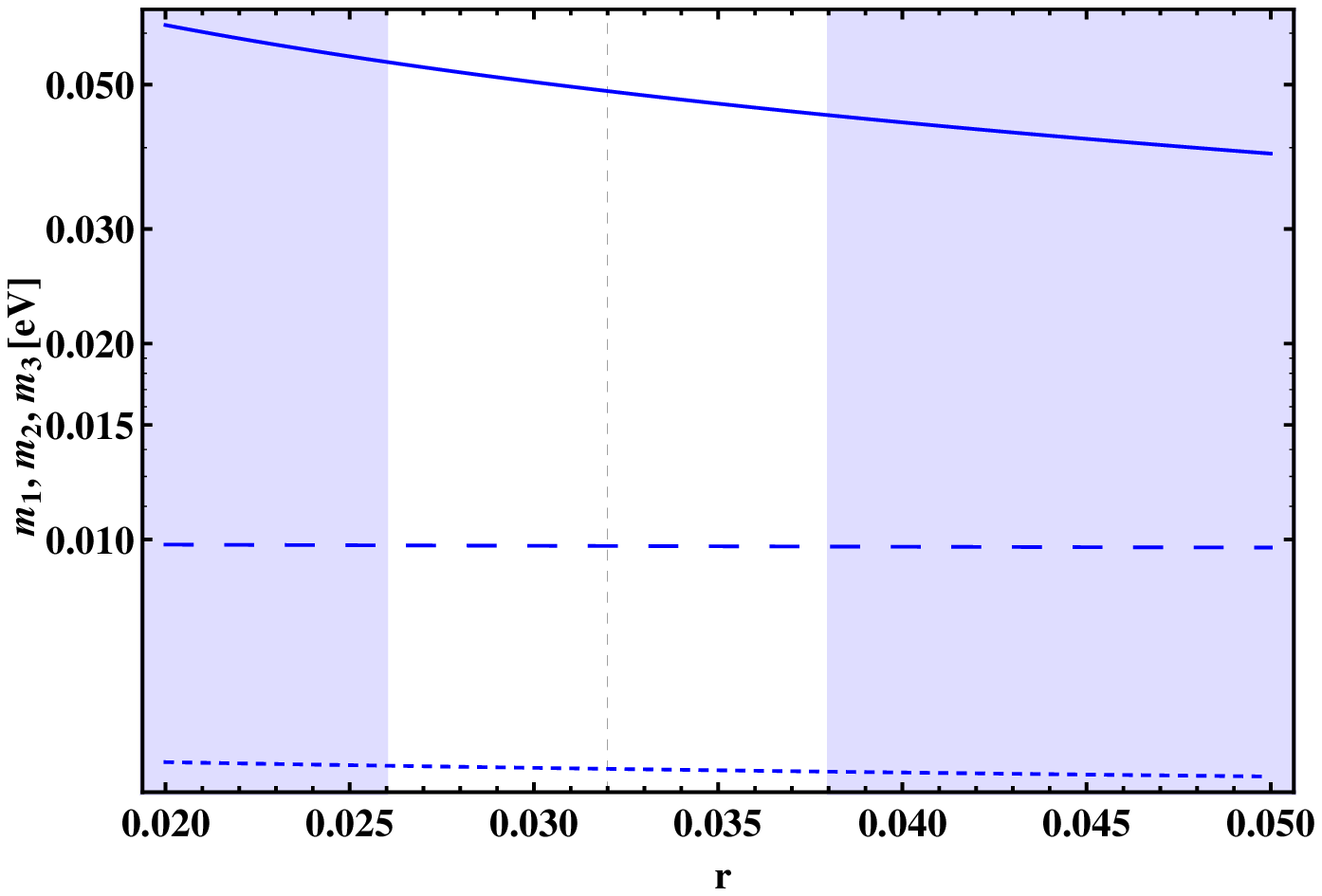}}
     \end{center}
   \caption{The values of the three light neutrino masses
corresponding to the solutions A (left panel) and B (right panel)
in the case of NO spectrum, versus $r$.
The dotted, dashed 
and solid lines  
correspond to the three
light neutrino masses
$m_1$, $m_2$, $m_3$.
    The gray region is excluded by present oscillation data. The
   vertical dashed line corresponds to the best fit value for
   $r=0.032$.
    See text for further details.
\label{fig:NOmasses}}
\end{figure}
%%%%%%%%%%%%%%%%%%%

%%%%%%%%%%%%%%%%%%%%%%%%%%%%%%%%%%%%%%
\begin{figure}
   \begin{center}
   {\includegraphics[width=7cm]{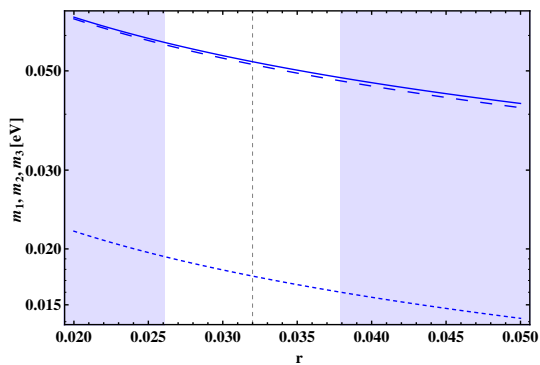}}
     \end{center}
   \caption{The values of the three
light neutrino masses in the case of the solution
corresponding to IO spectrum, versus $r$.
The dotted, dashed
and solid lines 
correspond to the three
light neutrino masses
$m_3$, $m_1$, $m_2$.
   The gray region is excluded by present oscillation data. The
   vertical dashed line corresponds to the best fit value for
   $r=0.032$.
\label{fig:IOmasses}}
\end{figure}
%%%%%%%%%%%%%%%%%%%

In the Figs.\ \ref{fig:NOmasses} and \ref{fig:IOmasses} we present the
dependence of the neutrino masses with respect to
$r$ for normal and inverted ordering respectively.

\subsection{The Mixing Angles and the Dirac and Majorana
CP Violation Phase}

 The PMNS neutrino mixing matrix received contributions from the
diagonalisation of the neutrino Majorana mass matrix $M_{\nu}$
and of the charged lepton mass matrix $M_e = v_dY_e$:
$U_{\text{PMNS}} = U_{eL}^{\dagger} U_{\nu}$, where  $U_{\nu}$
is given in eq.\ (\ref{Unu})
with $\tilde{Q}=\diag(1,e^{\ci \phi_2/2},e^{\ci \phi_3/2})$ and the values of the
phases $\phi_2$ and $\phi_3$ in the cases of NO and IO spectra
were specified in the preceding subsection.
The matrix of charged lepton Yukawa couplings $Y_e$,
eq.\ \eqref{eq:Yd}, and thus $M_e$, has a block-diagonal form.
The unitary matrix  $U_{eL}$ diagonalises the
the Hermitian matrix $M^{\dagger}_eM_e$:
$M^{\dagger}_eM_e = U_{eL} (M_e^{d})^2 U_{eL}^{\dagger}$,
where  $M_e^{d} = \text{diag}(m_e, m_\mu, m_\tau)$,
$m_l$ being the mass of the charged lepton $l$.
As a consequence of the block-diagonal form of
$M_e$, the matrix $U_{eL}$ can be
parametrised in terms of one mixing angle ($\theta^e_{12}$)
and one phase ($\varphi$):
$U_{eL} = \Phi \, R_{12}(\theta^e_{12})$, where $\Phi =
\text{diag}(1, \text{e}^{\ci \varphi},1)$ and
%%%%%%%%%%%%%%%%%%%%%%%%%%%%%%%%%%%%%
\begin{equation}
R_{12}(\theta^e_{12}) =
  \begin{pmatrix}
    \cos\theta^e_{12} & \sin\theta^e_{12} & 0 \\
    -\,\sin\theta^e_{12} &  \cos\theta^e_{12}  & 0 \\
   0 & 0 & 1 \\
  \end{pmatrix} \;.
\end{equation}
%%%%%%%%%%%%%%%%%%%%%%%%%%%%%%%%%%
%
Due to the $SU(5)$ symmetry of the model, $Y_d$ and $Y_e$
(and therefore the corresponding down quark and
charged lepton mass matrices)
are expressed in terms of the same parameters.
As a consequence, the angle $\theta^e_{12}$ in the model
considered is related to the Cabibbo angle $\theta^c \cong 0.226$.
Using, for example, the approximate formulas from
\cite{Marzocca:2011dh}, we find that
%%%%%%%%%%%%%%%%%%%%%%%%%%%%%%%%%
\begin{equation}
\theta^e_{12} \cong \left|\frac{b_d^{\prime}}{b_d}\right| \theta^c
\cong 0.9\,\theta^c\;,
\label{eq:the12}
\end{equation}
%%%%%%%%%%%%%%%%%%%%%%%%%%%%%%%
%
where we have used the values of $b_d^{\prime}$ and $b_d$ from
Table \ \ref{Tab:Parameters}.

 Comparing next the expressions on the two sides
sides of the equation
$M_e^{\dagger} M_e = U_{eL} (M_e^{d})^2 U_{eL}^\dagger$
we get, in particular:
%%%%%%%%%%%%%%%%%%%%%%%%
\begin{equation}
\text{e}^{- \ci (\varphi +\frac{\pi}{2})} \, (m_{\mu}^2 - m_e^2)\,
\cos \theta_{12}^e \sin \theta_{12}^e  = v_d^2
\left( \frac{3}{2} b_d a_d - 36 c_d b^{\prime}_d\right)\,.
\end{equation}

%%%%%%%%%%%%%%%%%%%%%
%
Using the fit results in Table\ \ref{Tab:Parameters} one
can check that the right hand side of the last equation is
real and positive. Comparing the phases of the two
expressions one concludes that
%%%%%%%%%%%%%%%%%%%%%%%%%
\begin{equation}
\varphi = \frac{3}{2}\pi\,.
\label{varphi}
\end{equation}
%%%%%%%%%%%%%%%%%%%%%%%%%%%%%%%%%%%%%%%%%%%%%%
%

In the approximation we are using the PMNS matrix is given  by:
%%%%%%%%%%%%%%%%%%%%%%%%%%%%%%%%%%%%
\begin{eqnarray}
\ti U_{\text{PMNS}}= \begin{pmatrix}
\sqrt{2/3} c_{12}^e + \sqrt{1/6}s_{12}^e e^{- \ci \varphi} &
\sqrt{1/3}c_{12}^e - \sqrt{1/3}s_{12}^e e^{- \ci \varphi}  &
\sqrt{1/2}s_{12}^e e^{- \ci \varphi}\\
 \sqrt{2/3}s^e_{12} - \sqrt{1/6}c^e_{12}\text{e}^{-\ci \varphi} &
\sqrt{1/3}s^e_{12} + \sqrt{1/3}c^e_{12}\text{e}^{-\ci \varphi}&
 -\sqrt{1/2} c^e_{12} \text{e}^{-\ci \varphi}\\
-\sqrt{1/6} & \sqrt{1/3} &  \sqrt{1/2}\\
\end{pmatrix} \,\tilde{Q}\;,
\label{UPMNS2}
\end{eqnarray}
%%%%%%%%%%%%%%%%%%%%%%%%%%%%%%%%%%%%%%%%%%%%%%%%
%
where $c^e_{12} = \cos\theta^e_{12}$,
$s^e_{12} = \sin\theta^e_{12}$ and
$\tilde{Q}$ is the diagonal phase matrix defined
in eq.\ (\ref{Unu}). It follows from the above expression
for the PMNS matrix that the angle $\theta_{13}$
is given approximately by
%%%%%%%%%%%%%%%%%%%%%%%%%%%%%%%%
\begin{equation}
\sin^2\theta_{13}\cong \frac{1}{2}C^2\, \sin^2\theta^c
\cong \frac{\sin^2\theta^c}{2.5} \cong 0.02\;,
~C \cong 0.9
\label{th13thC}
\end{equation}
%%%%%%%%%%%%%%%%%%%%%%%%%%%%%%
%
where we took into account the relation in eq.\ (\ref{eq:the12})
and the value of $C\equiv |b_d^{\prime}/b_d|$.

 As was shown in, e.g., \cite{Marzocca:2011dh},
the phase $\varphi$ and the Dirac phase
$\delta$ in eqs.\ (\ref{eq:UPMNS}) - (\ref{eq:V}) are related
(at leading order) as follows:
%%%%%%%%%%%%%%%%%%%%%%%%%%%%%%%%
\begin{equation}
\delta = \varphi + \pi\,.
\label{dtavarphi}
\end{equation}
%%%%%%%%%%%%%%%%%%%%%%%%%%%%%%
%
Thus, for the Dirac phase we get from (\ref{varphi}):
%%%%%%%%%%%%%%%%%%%%%%%%%%%%%%%%
\begin{equation}
\delta = \frac{\pi}{2}\,.
\end{equation}
%%%%%%%%%%%%%%%%%%%%%%%%%%%%%%
%

 Numerically, for $\varphi = 3\pi/2$ and $s^e_{12} = 0.203$
(see eq.\ \eqref{eq:the12}), the PMNS matrix, eq.\ \eqref{UPMNS2}, reads:
%%%%%%%%%%%%%%%%%%%%%%%%%%%%%%%%%%%%
\be
U_{\text{PMNS}} \cong \left(
\begin{array}{ccc}
0.804 e^{\ci 5.81^{\circ}} & 0.577 e^{-\ci 11.50^{\circ}}
& 0.144 e^{- \ci 270.000^{\circ}} \\
0.433 e^{-\ci 67.85^{\circ}}  & 0.577 e^{\ci 78.50^{\circ}} & -\,0.692 e^{-\ci 270.000^{\circ}}\\
-0.408   &  0.577 & 0.707
\end{array}
\right)\,\tilde{Q}\,.
\label{PMNSnumeric2}
\ee
%%%%%%%%%%%%%%%%%%%%%%%%%%%%%%%%%%
%
Thus, comparing the absolute values of the
elements $U_{e1}$, $U_{e2}$, $U_{\mu 3}$ and $U_{\tau 3}$
of the PMNS matrix in the standard parametrisation,
eq.\ \eqref{eq:UPMNS},
and in eq.\ \eqref{PMNSnumeric2},
we have: $c_{12}c_{13} = 0.804$,  $s_{12}c_{13} = 0.577$,
 $s_{23}c_{13} = 0.692$ and $c_{23}c_{13} = 0.707$.
Using the predicted value of $\theta_{13}$, eq.\ \eqref{th13thC},
these relations allow us to obtain the values of
$\theta_{12}$ and $\theta_{23}$.
We note that the tri-bimaximal mixing value of the
solar neutrino mixing angle $\theta_{12}$,
which corresponds to $\sin^2\theta_{12} = 1/3$,
is corrected by a quantity which, as it follows
from the general form of such corrections
\cite{FPR,HPR07,Marzocca:2011dh},
is determined by the angle $\theta_{13}$
and the Dirac phase $\delta$:
%%%%%%%%%%%%%%%%%%%%%%%%%
\be
\sin^2\theta_{12} \cong
\frac{1}{3} + \frac{2\sqrt{2}}{3}\,\sin\theta_{13}\,\cos\delta
\label{th120}
\ee
%
%%%%%%%%%%%%%%%%%%%%%%%%
%
where $\delta$ is the Dirac phase in the standard
parametrisation of the PMNS matrix.
As we have seen, to leading order $\delta = \pi/2$.
The Majorana phases  $\beta_1$, $\beta_2$ (or
$\alpha_{21}$ and $\alpha_{31}$) are determined,
as it follows from eqs.\ (\ref{eq:UPMNS}) and (\ref{UPMNS2})
(or (\ref{PMNSnumeric2})),
by the diagonal matrix $\tilde{Q}$ and
take CP conserving values.
Note, however, that the parametrisation
of the PMNS matrix in eq.\ (\ref{PMNSnumeric2})
differs from the standard one: it
corresponds to one of the several possible
parametrisations of the PMNS matrix \cite{HPR07}.
Thus, in order to get the values of the
Dirac and Majorana phases $\delta$ and
$\beta_1$, $\beta_2$
(or $\alpha_{21}$, $\alpha_{31}$),
of the standard parametrisation of the
PMNS matrix, one has to
bring the expressions (\ref{PMNSnumeric2})
in a form which corresponds to the ``standard'' one
in eq.\ (\ref{eq:UPMNS}). This can be done
by using the freedom of multiplying the rows of the
PMNS matrix with arbitrary phases and by shifting
some of the common phases of the columns to a diagonal
phase matrix $P$. The results for the
numerical matrix in eq.\ (\ref{PMNSnumeric2}) is:
%%%%%%%%%%%%%%%%%%%%%%%%%%%%%%%%%%%%
\be
U_{\text{PMNS}} \cong
\left(
\begin{array}{ccc}
0.804 & 0.577 & 0.144 e^{-\ci 84.25^{\circ}} \\
-\,0.433 e^{\ci 10.59^{\circ}}  & 0.577 e^{-\ci 5.75^{\circ}} & 0.692 \\
0.408 e^{-\ci 11.56^{\circ}}  & -\,0.577 e^{\ci 5.75^{\circ}}  & 0.707
\end{array}
\right)\,P\, \tilde{Q}\,,
\label{PMNSnumeric3}
\ee
where $\tilde{Q} = \diag(1,e^{\ci \phi_2/2},e^{\ci \phi_3/2}) =
e^{\ci \phi_3/2} \diag(e^{-\ci \phi_3/2},e^{-\ci(\phi_3-\phi_2)/2},1)$
and the new phase matrix
$P$ = $\diag(e^{\ci 11.50^{\circ}},e^{- \ci 5.81^{\circ}},-1)$.
Now comparing eq.\ (\ref{PMNSnumeric3})
with eq.\ (\ref{eq:UPMNS}) we can obtain the
values of the Dirac and the two Majorana phases
of the standard parametrisation of the
PMNS matrix, predicted by the model.
For the Dirac phase we find $\delta \cong 84.3^{\circ}$.
Note that the Majorana phases
$\beta_1/2$ and $\beta_2/2$
(or $\alpha_{21}/2$ and $\alpha_{31}/2$)
in the standard parametrisation
are not CP conserving \cite{Chen:2011vd}:
due to the matrix  $P$
they get CP violating corrections to the CP
conserving values 0 and $\pi/2$ or $3\pi/2$.

 As we have seen, the value of the Dirac
phase $\delta$ predicted by the model
is close to $\pi/2$. This implies
that the magnitude of the CP violation effects
in neutrino oscillations, is also predicted
to be be relatively large. Indeed,
the rephasing invariant associated
with the Dirac phase \cite{CJ85},
$J_{\text{CP}} = {\rm Im}(U^*_{e1}U_{\mu 1}U_{e3}U^*_{\mu3})$,
which  determines the magnitude of CP violation effects in
neutrino oscillations \cite{PKSP3nu88}, has the following
value:
%%%%%%%%%%%%%%%%%%%%%%%%%%
\be
J_{\text{CP}}= 0.0324\,.
\label{JCP}
\ee
%%%%%%%%%%%%%%%%%%%%%%%%%%%
%

 The values we have obtained for both $\sin\theta_{13}$ and $\delta$
are in very good agreement with the numerical
results in Table \ \ref{Tab:NeutrinoResults} derived using the {\tt
REAP} package \cite{Antusch:2005gp}.
%%%%%%%%%%%%%%%%%%%%%%%%%%%%%%%%%%%%%
\begin{table}
\begin{center}
\begin{tabular}{cccc}
\toprule
Quantity & Experiment (2$\sigma$ ranges) & Model  \\
\midrule
$\sin^2 \theta_{12}$ & 0.275 -- 0.342 & 0.340 &   \\
$\sin^2 \theta_{23}$ & 0.36 -- 0.60   & 0.490  &   \\
$\sin^2 \theta_{13}$ & 0.015 -- 0.032 & 0.020 &   \\
$\delta$      & - & 84.3$^\circ$ &   \\
\bottomrule
\end{tabular}
\caption{ \label{Tab:NeutrinoResults}
Numerical results for the neutrino sector. The experimental results
are taken from \protect\cite{Fogli:2011qn} apart from
the value for $\theta_{13}$ which is the DayaBay result
\cite{An:2012eh}.
}
\end{center}
\end{table}
%%%%%%%%%%%%%%%%%%%%%%%%%%%%%%%%%%%%%%%%%%
%

 It is possible to derive simple analytic expressions
which explain the numerical results obtained above and quoted in
Table \ref{Tab:NeutrinoResults}.
Indeed,  up to corrections of order $(\theta_{12}^e)^2$ we have:
%%%%%%%%%%%%%%%%%%%%
\begin{align}
\theta_{12} & =
\arcsin \frac{1}{\sqrt{3}} + \frac{\sqrt{2}}{8} (\theta_{12}^e)^2 \;, \\
\theta_{13} & = \frac{1}{\sqrt{2}} \theta_{12}^e \;,\\
\theta_{23} & = \frac{\pi}{4} - \frac{1}{4}(\theta_{12}^e)^2 \;,\\
\delta & = \frac{\pi}{2} - \frac{1}{2} \theta_{12}^e \;, \\
 \beta_1 & = 2 \pi - 2 \theta_{12}^e + \phi_3\;, \\
 \beta_2 & = 2 \pi + \theta_{12}^e + \phi_3 - \phi_2\;,
\label{AApprox}
\end{align}
%%%%%%%%%%%%%%%%%%%%%%%%%%%%%%%%%%%%%%%
%
where $\theta_{12}^e \cong 0.888 \theta^c$.
Note that the expression for $\delta$ is correct
up to $\mathcal{O}(\theta_{12}^e)$ only because it appears
always with $\theta_{13}$
which is of order $\theta_{12}^e$ itself.
Numerically, these approximations give for $\theta_{12}^e = 0.2$:
%%%%%%%%%%%%%%%%%%%%%%%%%%%
\begin{align}
\sin^2 \theta_{12} & = 0.340 \;, \\
\sin^2 \theta_{13} & = 0.020 \;, \\
\sin^2 \theta_{23} & = 0.490 \;, \\
\delta & = 84.3^\circ \;, \\
\beta_1 & = 337.1^\circ  + \phi_3\;, \\
\beta_2 & = 11.5^\circ + \phi_3 - \phi_2\;.
\label{NApprox}
\end{align}
%%%%%%%%%%%%%%%%%%%%%%%%%%
%
As we see, the results obtained using the approximate analytic
expressions are in very good agreement
with those derived in the numerical analysis.

Note that all these relations were derived neglecting RGE corrections.
Indeed they are under control. For the inverted ordering the RGE corrections
can be expected to be largest, because there $m_1$ and $m_2$ are almost equal
\cite{Antusch:2003kp}. We have found numerically with the REAP package
\cite{Antusch:2005gp} that the biggest deviation is in $\delta$ which goes
down to $81.2^\circ$. The Majorana phases run less than one degree and also
the mixing angles stay well within their two sigma ranges.

\subsection{Predictions for Other Observables in the Neutrino Sector}

 We derive in this section the predictions for the sum of the
neutrino masses and the effective Majorana mass $\meff$
in neutrinoless double beta decay (see, e.g., \cite{BiPet87})
using the standard parameterisation of the
PMNS mixing matrix as in \eqref{eq:V} and the results on the
neutrino masses, mixing angles and CPV phases obtained in
preceding subsections of this Section.

  In the case of solution A for the NO neutrino mass spectrum
we get for the sum of the neutrino masses:
%%%%%%%%%%%%%%%%%%%%%%%%%%%%%%%%%%%
\begin{equation}
\sum_{k=1}^3 m_k=  6.31 \times 10^{-2}~\text{eV} \;,~~{\rm solution~A~(NO)}\,.
\label{summkA}
\end{equation}
%%%%%%%%%%%%%%%%%%%%%%%%%%%%%%%%
%
In this case we have $\phi_2 = \phi_3 = 0$ (see subsection 4.1)
and for the effective Majorana mass we obtain
using eqs.\ (\ref{eq:m123A}) and (\ref{PMNSnumeric3}):
%%%%%%%%%%%%%%%%%%%%%%%%%%%%%%%
\be
\meff = |\sum_{k=1}^3 (U_{\text{PMNS}})^2_{ek} m_k|=
4.90\times 10^{-3}~\text{eV}\;,~~{\rm solution~A~(NO)}\,.
\label{meffA}
\ee
%%%%%%%%%%%%%%%%%%%%%%%%%%%%%%%%
%
The same quantities for solution B of the NO spectrum
have the values:
%%%%%%%%%%%%%%%%%%%%%%%%%%%%%%%%%%%
\begin{equation}
\sum_{k=1}^3 m_k=  6.54 \times 10^{-2}~\text{eV} \;,~~{\rm solution~B~(NO)}\,,
\label{summkB}
\end{equation}
%%%%%%%%%%%%%%%%%%%%%%%%%%%%%%%%
%
and
%%%%%%%%%%%%%%%%%%%%%%%%%%%%%%%
\be
\meff =
7.95\times 10^{-3}~\text{eV}\;,~~{\rm solution~B~(NO)}\,,
\label{meffB}
\ee
%%%%%%%%%%%%%%%%%%%%%%%%%%%%%%%%
%
where we have used the fact that for
solution B we have $\phi_2 = 0$ and $\phi_3 = \pi$.
As a consequence, in particular, of the values of
$\phi_{2,3}$, the three terms in the expression for
$\meff$ essentially add.

 Finally, in the case of IO spectrum we obtain:
%%%%%%%%%%%%%%%%%%%%%%%%%%%%%%%%%%%
\begin{equation}
\sum_{k=1}^3 m_k=  12.1 \times 10^{-2}~\text{eV} \;,~~{\rm (IO)}\,,
\label{summkIO}
\end{equation}
%%%%%%%%%%%%%%%%%%%%%%%%%%%%%%%%
%
and
%%%%%%%%%%%%%%%%%%%%%%%%%%%%%%%
\be
\meff =
2.17\times 10^{-2}~\text{eV}\;,~~{\rm (IO)}\,,
\label{meffIO}
\ee
%%%%%%%%%%%%%%%%%%%%%%%%%%%%%%%%
%
We recall that for the IO spectrum
we have $\phi_2 = \pi$ and $\phi_3 = 0$
and there is a partial compensation
in $\meff$ between the dominant
contributions due to the terms
$\propto m_1$ and $\propto m_2$.

\section{Summary and Conclusions}

We have presented here
the first $SU(5) \times T^{\prime}$
unified model of flavour,
which predicts the reactor neutrino mixing angle
$\theta_{13}$ to be in the range determined
 by DayaBay \cite{An:2012eh} and RENO
\cite{RENO0412} experiments, and all other mixing
angles are predicted to have values within the experimental
uncertainties. It implements a type I seesaw mechanism and from the breaking
of the discrete family symmetry $T^{\prime}$ we obtain tri-bimaximal
mixing in the neutrino sector. The relatively large value of
$\theta_{13}$ is then generated entirely by corrections coming
from the charged lepton sector. This is a generic effect in GUTs
where Yukawa couplings are related to each other. Here we
have used recently proposed $SU(5)$ GUT relations \cite{Antusch:2009gu}
between the down-type quark Yukawa matrix
and the charged lepton Yukawa matrix to get the relatively large prediction
for the reactor mixing angle
$\theta_{13}$ along the lines
proposed in \cite{Marzocca:2011dh,Antusch:2011qg}.

The corrections to the solar and the atmospheric neutrino mixing angle
are under control due to the structure of the charged lepton Yukawa matrix
and the pattern of the complex CP violation phases.
The model exhibits a special kind of
CP ciolation, the so-called ``geometrical'' CP violation.
All parameters and vevs are real and all non-trivial
phases are coming from the complex Clebsch--Gordan
coefficients of $T^{\prime}$
and are integer multiples of $\pi/4$.
We have given the renormalisable superpotential which
generates
effectively the Yukawa matrices after integrating out heavy messenger
fields and plugging in the family symmetry breaking flavon vevs
which was missing so far in the literature for
$SU(5) \times T^{\prime}$ models.
The flavon vevs point in special directions in flavour space
and are all real. These results come out as solutions
to the flavon alignment superpotential we have presented
in the Appendix.

We have shown, in particular, that the
phase pattern in the Yukawa matrices actually gives a good fit
of the quark and charged lepton masses and the CKM parameters at low
energies. This fit fixes the charged lepton Yukawa matrix completely
and since we find tri-bimaximal mixing in the neutrino sector itself,
we can make predictions for the neutrino masses and all PMNS
parameters.
The angle $\theta_{13}$ is predicted to have a value
corresponding to $\sin^2\theta_{13} \cong
0.8\,\sin^2\theta^c/2 = 0.02$. For the Dirac phase $\delta$ we
obtain in the standard parametrisation of the PMNS matrix $\delta =
84.3^{\circ}$.
Our model also predicts $\sin^2\theta_{12} = 0.340$ and
$\sin^2\theta_{23} = 0.490$.
There are three different possible solutions for the neutrino masses,
two with normal ordering (NO, solutions A and B)
and one with inverted ordering (IO).
All three cases can be tested in experiments determining the
absolute neutrino mass scale (or the sum of the three neutrino masses),
in experiments which can measure the solar and atmospheric neutrino
mixing angles with a high precision, in experiments
searching for CP violation in neutrino oscillations and in
neutrinoless double beta decay experiments.
For the sum of three neutrino masses we get (with relatively small
uncertainties, see Figs. (2) and (3)):
$\sum_{k=1}^3 m_k= 6.31 \times 10^{-2}~\text{eV}$ (NO, A);
$6.54 \times 10^{-2}~\text{eV}$ (NO, B) and
$12.1 \times 10^{-2}~\text{eV}$ (IO).
The $\betabeta$-decay effective Majorana mass for the three solutions
is also unambiguously predicted:
$\meff =  4.90 \times 10^{-3}$~eV (NO, A);
$7.95 \times 10^{-3}$~eV (NO, B);
$2.17 \times 10^{-2}$~eV (IO).
The three solutions differ only in the values of
the three neutrino masses and of the Majorana phases,
so that we make one single prediction for the
rephasing invariant which determines the magnitude of CP violation
effects in neutrino oscillations: $J_{\text{CP}} = 0.0324$.
This value of $J_{\text{CP}}$ is relatively large
and can be tested in the experiments
on CP violation in neutrino oscillations.

In conclusion, with the recent measurement of the
last unknown neutrino mixing angle,
neutrino physics has entered a new era.
All angles are determined with a rather good precision,
constraining flavour models severely.
Since $\theta_{13}$ turned out to be relatively
large, the observation of CP violation in the lepton
sector might be feasible with data from the running
and upcoming neutrino oscillation experiments.
Explaining the data on leptonic CP violation
would pose another challenge for flavour models.
The model we proposed here is from this point of view
rather comprehensive combining many ideas which
have been proposed elsewhere but have been combined here
consistently for the first time.
Due to the GUT structure
we can fit the quark masses and mixing parameters and the
charged lepton masses, and using the latter we make
definite predictions for the neutrino mass spectrum, the leptonic
mixing angles and the leptonic CP violating phases.
Our model is therefore testable in a variety of experiments.
We are looking forward to the outcome of these tests.

\section*{Acknowledgements}
We would like to thank Christoph Luhn and Ferruccio Feruglio for useful discussions.
This work was supported in part by the INFN program on
``Astroparticle Physics'', by the Italian MIUR program on
``Neutrinos, Dark Matter and  Dark Energy in the Era of LHC''
(A.M. and S.T.P.) and by the World Premier International
Research Center Initiative (WPI Initiative), MEXT,
Japan  (S.T.P.).
Furthermore the authors acknowledge partial support from the European Union
under FP7 ITN INVISIBLES (Marie Curie Actions,
PITN-GA-2011-289442).

\appendix

\section{$\boldsymbol{T^{\prime}}$: The Rules of the Game} \label{App:Tprime}

\begin{table}
\centering
\renewcommand{\tabcolsep}{1.1mm}
{\footnotesize\begin{tabular}{c} \toprule
\\
$a\otimes \Gamma^p =  a\Gamma^p$, \quad $a\otimes a'(a'')= a'(a'')$,
\quad $a'\otimes a'(a'')= a'' (a)$, $\quad$ $a'(a'') \otimes a''= a (a')$  \\
\\
$\begin{pmatrix}
x_1  \\
x_2 \end{pmatrix}_\db \otimes a' (a'') =\begin{pmatrix}
x_1 a'(a'')\\
x_2 a'(a'')\end{pmatrix}_{\db'(\db'')} $, \quad $\begin{pmatrix}
y_1  \\
y_2  \end{pmatrix}_{\db'} \otimes a' (a'') =\begin{pmatrix}
y_1 a'(a'')  \\
y_2 a'(a'')  \end{pmatrix}_{\db''(\db)} $,
$\begin{pmatrix}
 z_1\\
 z_2\end{pmatrix}_{\db''} \otimes a' (a'') = \begin{pmatrix}
 z_1 a'(a'')  \\
 z_2 a'(a'')  \end{pmatrix}_{\db(\db')} $\\
\\
$\begin{pmatrix}
x_1 \\
x_2 \end{pmatrix}_{\db (\db')}\otimes \begin{pmatrix}
x_1' \\
x_2' \end{pmatrix}_{\db (\db'')} = \left( \dfrac{x_1 x_2'- x_2
x_1'}{\sqrt{2}} \right)_\si  \oplus \left( \begin{array}{c}
\frac{(1-\ci)}{2}(x_1 x_2'+ x_2 x_1') \\
  \ci x_1 x_1' \\
 x_2 x_2'  \end{array} \right)_\tp $\\
\\
$\begin{pmatrix}
y_1 \\
y_2 \end{pmatrix}_{\db' (\db)}\otimes \begin{pmatrix}
y_1' \\
y_2' \end{pmatrix}_{\db '(\db'')} = \left( \dfrac{y_1 y_2'- y_2
y_1'}{\sqrt{2}} \right)_{\si''} \oplus \left( \begin{array}{c}
\ci y_1 y_1' \\
 y_2 y_2'  \\
\frac{(1-\ci)}{2}(y_1 y_2'+ y_2 y_1')  \end{array} \right)_\tp $\\
\\
$\begin{pmatrix}
z_1 \\
z_2 \end{pmatrix}_{\db'' (\db)}\otimes \begin{pmatrix}
z_1' \\
z_2' \end{pmatrix}_{\db''(\db')} = \left( \dfrac{z_1 z_2'- z_2
z_1'}{\sqrt{2}} \right)_{\si'}  \oplus \left( \begin{array}{c}
z_2 z_2' \\
\frac{(1-\ci)}{2}(z_1 z_2'+ z_2 z_1') \\
\ci z_1 z_1'  \end{array} \right)_\tp $\\ \\
$(a')_{\si'}\otimes \left( \begin{array}{c}
u_1 \\
u_2 \\
u_3 \end{array} \right)_\tp= \left( \begin{array}{c}
u_3 a'  \\
u_1  a' \\
u_2  a' \end{array} \right)_\tp $, \quad $(a'')_{\si''}\otimes
\left( \begin{array}{c}
u_1 \\
u_2 \\
u_3 \end{array} \right)_\tp= \left( \begin{array}{c}
u_2 a''  \\
u_3  a'' \\
u_1  a'' \end{array} \right)_\tp $\\ \\
$\begin{pmatrix}
x_1 \\
x_2 \end{pmatrix}_{\db}\otimes \left( \begin{array}{c}
u_1 \\
u_2 \\
u_3 \end{array} \right)_\tp= \frac{1}{\sqrt{3}}\left[
\begin{pmatrix}
(1+\ci) x_2 u_2 +x_1 u_1\\
(1-\ci) x_1u_3 -x_2 u_1 \end{pmatrix}_{\db}  \oplus  \begin{pmatrix}
(1+\ci) x_2 u_3 +x_1 u_2\\
(1-\ci) x_1u_1 -x_2 u_2 \end{pmatrix}_{\db'} \oplus    \begin{pmatrix}
(1+\ci) x_2 u_1+ x_1 u_3\\
(1-\ci) x_1u_2 -x_2 u_3 \end{pmatrix}_{\db''}   \right] $\\ \\
$\begin{pmatrix}
y_1 \\
y_2 \end{pmatrix}_{\db'}\otimes \left( \begin{array}{c}
u_1 \\
u_2 \\
u_3 \end{array} \right)_\tp= \frac{1}{\sqrt{3}}\left[
\begin{pmatrix}
(1+\ci)y_2 u_1+ y_1 u_3 \\
(1-\ci) y_1u_2 -y_2 u_3 \end{pmatrix}_{\db}  \oplus
\begin{pmatrix}
(1+\ci) y_2 u_2 +y_1 u_1\\
(1-\ci) y_1u_3 -y_2 u_1 \end{pmatrix}_{\db'}  \oplus  \begin{pmatrix}
(1+\ci) y_2 u_3 +y_1 u_2\\
(1-\ci) y_1u_1 -y_2 u_2 \end{pmatrix}_{\db''}   \right] $\\ \\
$\begin{pmatrix}
z_1 \\
z_2 \end{pmatrix}_{\db''}\otimes \left( \begin{array}{c}
u_1 \\
u_2 \\
u_3 \end{array} \right)_\tp= \frac{1}{\sqrt{3}}\left[
\begin{pmatrix}
(1+\ci) z_2 u_3+z_1 u_2\\
(1-\ci) z_1u_1 -z_2 u_2 \end{pmatrix}_{\db} \oplus  \begin{pmatrix}
(1+\ci) z_2 u_1 +z_1 u_3\\
(1-\ci) z_1u_2 -z_2 u_3 \end{pmatrix}_{\db'}  \oplus
\begin{pmatrix}
(1+\ci) z_2 u_2 +z_1 u_1\\
(1-\ci) z_1u_3 -z_2 u_1 \end{pmatrix}_{\db''}     \right] $\\ \\
$\left( \begin{array}{c}
u_1 \\
u_2 \\
u_3 \end{array} \right)_\tp \otimes \left( \begin{array}{c}
u_1' \\
u_2' \\
u_3' \end{array} \right)_\tp= \frac{1}{\sqrt{3}}\left[ (u_1 u_1' +
u_2 u_3' + u_3 u_2')_{\si} \oplus  (u_1 u_2' + u_2 u_1' + u_3
u_3')_{\si'}
\oplus  (u_1 u_3' + u_2 u_2' + u_3 u_1')_{\si''}\right]\oplus $\\
$\oplus \frac{1}{\sqrt{6}}\left( \begin{array}{c}
2u_1 u_1' - u_2 u_3' - u_3 u_2' \\
2u_3 u_3' - u_1 u_2' - u_2 u_1' \\
2u_2 u_2' - u_1 u_3' - u_3 u_1' \end{array} \right)_\tp \oplus
\frac{1}{\sqrt{2}}\left( \begin{array}{c}
 u_2 u_3' - u_3 u_2' \\
 u_1 u_2' - u_2 u_1' \\
 u_3 u_1' - u_1 u_3' \end{array} \right)_\tp$\\
\\
\bottomrule
\end{tabular}} \caption{\label{tab:CGs} The
Clebsch--Gordan coefficients for the tensor products
of $T^{\prime}$.}
\end{table}

$T^{\prime}$ is the double-covering group of the tetrahedral
symmetry $T$ which is isomorphic to $A_4$, the group of the even
permutations of four objects.  $T^{\prime}$ contains three
inequivalent one-dimensional representations, called $\mathbf{1}$,
$\mathbf{1}^{\prime}$ and $\mathbf{1}^{\prime\prime}$, one
three-dimensional, $\mathbf{3}$ and three two-dimensional
representations, $\mathbf{2}$, $\mathbf{2}^{\prime}$
and $\mathbf{2}^{\prime\prime}$. Two of these representations are
real, $\mathbf{1}$ and $\mathbf{3}$, one is pseudo-real $\mathbf{2}$
and the other four ones are complex. We list in Tab.\
\ref{tab:CGs} the relevant tensor products of $T^{\prime}$.
For more details on $T^{\prime}$, see, e.g.
\cite{Feruglio:2007uu} and references therein.

\section{Messenger Sector} \label{App:Messengers}

\begin{table}
\centering
\begin{tabular}{c ccc ccccccc}
\toprule
Messenger Fields & $SU(5)$ & $T^{\prime}$ & $U(1)_R$ &  $Z_{12}^u$ &   $Z_{8}^d$ &  $Z_{8}^\nu $ & $Z_{8} $ & $Z_{6} $ & $Z_{6}^{\prime} $&  $Z_{4} $\\
\midrule
$\Sigma_{1}^a$, $\bar{\Sigma}_{1}^a$ & $\mathbf{1}$, $\mathbf{1}$ & $\mathbf 1$, $\mathbf 1$ & 0, 2  & 4, 8 & 0, 0 &  0, 0  & 0, 0 & 2, 4 & 2, 4 & 0, 0\\
$\Sigma_{1}^b$, $\bar{\Sigma}_{1}^b$ & $\mathbf{1}$, $\mathbf{1}$ & $\mathbf 1$, $\mathbf 1$ & 0, 2  & 4, 8 & 0, 0 &  0, 0  & 0, 0 & 0, 0 & 0, 0 & 0, 0\\
$\Sigma_{1^{\prime}}^a$, $\bar{\Sigma}_{1^{\prime \prime}}^a$ & $\mathbf{1}$, $\mathbf{1}$ & $\mathbf 1^{\prime}$, $\mathbf 1^{\prime \prime}$ & 0, 2 &  6, 6 & 4, 4, &  4, 4 & 4, 4 & 0, 0  &  0, 0 & 2, 2\\
$\Sigma_{1^{\prime}}^b$, $\bar{\Sigma}_{1^{\prime \prime}}^b$ & $\mathbf{1}$, $\mathbf{1}$ & $\mathbf 1^{\prime}$, $\mathbf 1^{\prime \prime}$ & 0, 2 & 8, 4 & 4, 4, &  4, 4 & 4, 4 & 4, 2 & 4, 2 & 0, 0\\
$\Sigma_{1^{\prime}}^c$, $\bar{\Sigma}_{1^{\prime \prime}}^c$ & $\mathbf{1}$, $\mathbf{1}$ & $\mathbf 1^{\prime}$, $\mathbf 1^{\prime \prime}$ & 0, 2 & 8, 4 & 0, 0 & 0, 0 & 0, 0 & 2, 4 & 2, 4 & 0, 0\\
$\Sigma_{1^{\prime \prime}}^a$, $\bar{\Sigma}_{1^{\prime}}^a$ & $\mathbf{1}$, $\mathbf{1}$ & $\mathbf 1^{\prime \prime}$, $\mathbf 1^{\prime}$ & 0, 2 & 6, 6 & 4, 4 & 0, 0 & 4, 4 & 1, 5 & 1, 5 & 0, 0\\
$\Sigma_{1^{\prime \prime}}^b$, $\bar{\Sigma}_{1^{\prime}}^b$ & $\mathbf{1}$, $\mathbf{1}$ & $\mathbf 1^{\prime \prime}$, $\mathbf 1^{\prime}$ & 0, 2 &  0, 0 & 6, 2 & 2, 6 &  6, 2 & 3, 3 & 3, 3 & 0, 0\\
$\Sigma_{1^{\prime \prime}}^c$, $\bar{\Sigma}_{1^{\prime}}^c$ & $\mathbf{24}$, $\mathbf{24}$ & $\mathbf 1^{\prime \prime}$, $\mathbf 1^{\prime}$ & 0, 2 & 3, 9 & 2, 6 &  0, 0 & 6, 2 & 0, 0 &  3, 3 & 1, 3\\
$\Sigma_{2^{\prime \prime}}^a$, $\bar{\Sigma}_{2^{\prime}}^a$ & $\mathbf{1}$, $\mathbf{1}$ & $\mathbf 2^{\prime \prime}$, $\mathbf 2^{\prime}$ & 0, 2 & 9, 3 & 5, 3 & 7, 1 & 1, 7 & 3, 3 & 3, 3 & 3, 1 \\
$\Sigma_{2^{\prime \prime}}^b$, $\bar{\Sigma}_{2^{\prime}}^b$ & $\mathbf{1}$, $\mathbf{1}$ & $\mathbf 2^{\prime \prime}$, $\mathbf 2^{\prime}$ & 0, 2 & 9, 3 & 4, 4 & 5, 3 & 3, 5 & 4, 2 & 1, 5 & 2, 2\\
$\Sigma_{3}^a$, $\bar{\Sigma}_{3}^a$ & $\mathbf{1}$, $\mathbf{1}$ & $\mathbf 3$, $\mathbf 3$ & 0, 2 &  6, 6 & 4, 4 & 0, 0 & 4, 4 & 1, 5 & 1, 5 & 0, 0\\
$\Sigma_{3}^b$, $\bar{\Sigma}_{3}^b$ & $\mathbf{1}$, $\mathbf{1}$ & $\mathbf 3$, $\mathbf 3$ & 0, 2 &  0, 0 & 6, 2 & 2, 6 & 6, 2 & 3, 3 & 3, 3 & 0, 0\\
$\Sigma_{3}^c$, $\bar{\Sigma}_{3}^c$ & $\mathbf{1}$, $\mathbf{1}$ & $\mathbf 3$, $\mathbf 3$ & 0, 2 &  0, 0 & 0, 0 & 4, 4 & 0, 0 & 3, 3 & 3, 3 & 2, 2\\
$\Sigma_{3}^d$, $\bar{\Sigma}_{3}^d$ & $\mathbf{1}$, $\mathbf{1}$ & $\mathbf 3$, $\mathbf 3$ & 0, 2 &  0, 0 & 0, 0 & 0, 0 & 4, 4 & 0, 0 & 3, 3 & 2, 2\\
\midrule
$\Xi_{1^{\prime}}^a$, $\bar{\Xi}_{1^{\prime \prime}}^a$ & $\mathbf{5}$, $\bar{\mathbf{5}}$ & $\mathbf 1^{\prime}$, $\mathbf 1^{\prime \prime}$ & 1, 1 & 5, 7 & 0, 0 & 4, 4 & 0, 0 & 5, 1 & 5, 1 & 2, 2\\
$\Xi_{2^{\prime}}^a$, $\bar{\Xi}_{2^{\prime \prime}}^a$ & $\mathbf{5}$, $\bar{\mathbf{5}}$ & $\mathbf 2^{\prime}$, $\mathbf 2^{\prime \prime}$ & 1, 1 & 2, 10 & 7, 1 & 5, 3 & 3, 5 & 2, 4 & 5, 1 & 3, 1 \\
$\Xi_{2^{\prime \prime}}^a$, $\bar{\Xi}_{2^{\prime}}^a$ & $\mathbf{5}$, $\bar{\mathbf{5}}$ & $\mathbf 2^{\prime \prime}$, $\mathbf 2^{\prime}$ & 1, 1 & 8, 4 & 5, 3 & 7, 1 & 1, 7 & 2, 4 & 5, 1 & 1, 3\\
\midrule
$\Omega_{1}^a$, $\bar{\Omega}_{1}^a$ & $\mathbf{5}$, $\bar{\mathbf{5}}$ & $\mathbf 1$, $\mathbf 1$ & 0, 2  & 2, 10 & 0, 0 & 2, 10 & 4, 4 & 5, 1 &  5, 1 & 0,0\\
$\Omega_{1^{\prime}}^a$, $\bar{\Omega}_{1^{\prime \prime}}^a$ & $\mathbf{5}$, $\bar{\mathbf{5}}$ & $\mathbf 1^{\prime}$, $\mathbf 1^{\prime \prime}$ & 2, 0 & 8, 4 & 0, 0 & 8, 4 & 0, 0 & 4, 2 & 4, 2 & 2, 2\\
$\Omega_{1^{\prime}}^b$, $\bar{\Omega}_{1^{\prime \prime}}^b$ & $\mathbf{5}$, $\bar{\mathbf{5}}$ & $\mathbf 1^{\prime}$, $\mathbf 1^{\prime \prime}$ & 2, 0 & 9, 3 & 1, 7 & 9, 3 & 2, 6 & 4, 2 & 1, 5 & 2, 2\\
$\Omega_{2^{\prime \prime}}^a$, $\bar{\Omega}_{2^{\prime}}^a$ & $\mathbf{5}$, $\bar{\mathbf{5}}$ & $\mathbf 2^{\prime \prime}$, $\mathbf 2^{\prime}$ & 2, 0 & 9, 3 & 2, 6 & 9, 3 & 7, 1 & 1, 5 & 4, 2 & 2, 2\\
$\Omega_{3}^a$, $\bar{\Omega}_{3}^a$ & $\mathbf{5}$, $\bar{\mathbf{5}}$ & $\mathbf 1$, $\mathbf 1$ & 0, 2 & 0, 0 & 3, 5 & 0, 0 & 4, 4 & 4, 2 & 4, 2 & 1, 3 \\
\midrule
$\Upsilon_{1^{\prime \prime}}^a$, $\bar{\Upsilon}_{1^{\prime}}^a$ & $\mathbf{10}$, $\bar{\mathbf{10}}$ & $\mathbf 1^{\prime \prime}$, $\mathbf 1^{\prime}$ & 1, 1 & 2, 10 & 1, 7 & 5, 3 & 2, 6 & 3, 3  & 3, 3 & 2, 2 \\
$\Upsilon_{1^{\prime \prime}}^b$, $\bar{\Upsilon}_{1^{\prime}}^b$ & $\mathbf{10}$, $\bar{\mathbf{10}}$ & $\mathbf 1^{\prime \prime}$, $\mathbf 1^{\prime}$ & 1, 1 & 2, 10 & 0, 0 & 3, 5 & 4, 4 & 5, 1 & 2, 4 & 3, 1 \\
$\Upsilon_{2}^a$, $\bar{\Upsilon}_{2}^a$ & $\mathbf{10}$, $\bar{\mathbf{10}}$ & $\mathbf 2$, $\mathbf 2$ & 1, 1 & 11, 1 & 0, 0 & 2, 6 & 1, 7 &  4, 2 & 1, 5 & 3, 1 \\
$\Upsilon_{2}^b$, $\bar{\Upsilon}_{2}^b$ & $\mathbf{10}$, $\bar{\mathbf{10}}$ & $\mathbf 2$, $\mathbf 2$ & 1, 1 & 5, 7 & 2, 6 & 0, 0 & 7, 1 &  0, 0 &  0, 0 & 3, 1 \\
$\Upsilon_{2}^c$, $\bar{\Upsilon}_{2}^c$ & $\mathbf{10}$, $\bar{\mathbf{10}}$ & $\mathbf 2$, $\mathbf 2$ & 1, 1 & 5, 7 & 6, 2 & 4, 4 & 3, 5 &  0, 0& 0, 0 &  3, 1\\
$\Upsilon_{2^{\prime}}^a$, $\bar{\Upsilon}_{2^{\prime \prime}}^a$ & $\mathbf{10}$, $\bar{\mathbf{10}}$ & $\mathbf 2^{\prime}$, $\mathbf 2^{\prime \prime}$ & 1, 1 & 11, 1 & 0, 0 & 2, 6 & 1, 7 & 4, 2 & 1, 5 & 3, 1 \\
$\Upsilon_{2^{\prime}}^b$, $\bar{\Upsilon}_{2^{\prime \prime}}^b$ & $\mathbf{10}$, $\bar{\mathbf{10}}$ & $\mathbf 2^{\prime}$, $\mathbf 2^{\prime \prime}$ & 1, 1 & 5, 7 &  0, 0 & 4, 4 & 7, 1 & 4, 2 & 1, 5 & 3, 1 \\
$\Upsilon_{2^{\prime}}^c$, $\bar{\Upsilon}_{2^{\prime \prime}}^c$ & $\mathbf{10}$, $\bar{\mathbf{10}}$ & $\mathbf 2^{\prime}$, $\mathbf 2^{\prime \prime}$ & 1, 1 & 5, 7 & 2, 6 & 0, 0 & 7, 1 &   0, 0 & 0, 0 & 3, 1 \\
$\Upsilon_{2^{\prime}}^d$, $\bar{\Upsilon}_{2^{\prime \prime}}^d$ & $\mathbf{10}$, $\bar{\mathbf{10}}$ & $\mathbf 2^{\prime}$, $\mathbf 2^{\prime \prime}$ & 1, 1 & 11, 1 &  0, 0 & 2, 6 & 5, 3 & 2, 4 & 5, 1 & 3, 1 \\
$\Upsilon_{2^{\prime \prime}}^a$, $\bar{\Upsilon}_{2^{\prime}}^a$ & $\mathbf{10}$, $\bar{\mathbf{10}}$ & $\mathbf 2^{\prime \prime}$, $\mathbf 2^{\prime}$ & 1, 1 &  5, 7 & 4, 4 & 0, 0 & 7, 1 & 3, 3 & 0, 0 & 1, 3 \\
$\Upsilon_{2^{\prime \prime}}^b$, $\bar{\Upsilon}_{2^{\prime}}^b$ & $\mathbf{10}$, $\bar{\mathbf{10}}$ & $\mathbf 2^{\prime \prime}$, $\mathbf 2^{\prime}$ & 1, 1 & 11, 1 & 0, 0 & 2, 6 &  1, 7 & 4, 2 & 1, 5 &  3, 1 \\
$\Upsilon_{2^{\prime \prime}}^c$, $\bar{\Upsilon}_{2^{\prime}}^c$ & $\mathbf{10}$, $\bar{\mathbf{10}}$ & $\mathbf 2^{\prime \prime}$, $\mathbf 2^{\prime}$ & 1, 1 & 2, 10 & 2, 6 & 6, 2 & 7, 1 & 0, 0 & 3, 3 &  0, 0\\
$\Upsilon_{2^{\prime \prime}}^d$, $\bar{\Upsilon}_{2^{\prime}}^d$ & $\mathbf{10}$, $\bar{\mathbf{10}}$ & $\mathbf 2^{\prime \prime}$, $\mathbf 2^{\prime}$ & 1, 1 & 11, 1 & 0, 0 & 2, 6 &  5, 3  & 2, 4 & 5, 1 & 3, 1 \\
$\Upsilon_{2^{\prime \prime}}^e$, $\bar{\Upsilon}_{2^{\prime}}^e$ & $\mathbf{10}$, $\bar{\mathbf{10}}$ & $\mathbf 2^{\prime \prime}$, $\mathbf 2^{\prime}$ & 1, 1 & 11, 1 & 0, 0 & 6, 2 & 5, 3   &  0, 0 & 0, 0 &  1, 3\\
$\Upsilon_{3}^a$, $\bar{\Upsilon}_{3}^a$ & $\mathbf{10}$, $\bar{\mathbf{10}}$ & $\mathbf 3$, $\mathbf 3$ & 1, 1 & 2, 10 & 0, 0 & 7, 1 &  4, 4 & 4, 2 & 1, 5 &  1, 3\\
$\Upsilon_{3}^b$, $\bar{\Upsilon}_{3}^b$ & $\mathbf{10}$, $\bar{\mathbf{10}}$ & $\mathbf 3$, $\mathbf 3$ & 1, 1 & 8, 4 & 0, 0 & 5, 3 & 2, 6 &  2, 4 &  5, 1 & 1, 3 \\
$\Upsilon_{3}^c$, $\bar{\Upsilon}_{3}^c$ & $\mathbf{10}$, $\bar{\mathbf{10}}$ & $\mathbf 3$, $\mathbf 3$ & 1, 1 & 8, 4 & 2, 6 & 5, 3 & 6, 2 &  5, 1 & 5, 1 &  3, 1\\
$\Upsilon_{3}^d$, $\bar{\Upsilon}_{3}^d$ & $\mathbf{10}$, $\bar{\mathbf{10}}$ & $\mathbf 3$, $\mathbf 3$ & 1, 1 & 2, 10 & 5, 3 & 5, 3 & 6, 2 &  3, 3 & 0, 0 &  0, 0\\
\midrule
$\Gamma_{2^{\prime \prime}}^a$, $\bar{\Gamma}_{2^{\prime}}^a$ & $\mathbf{24}$, $\mathbf{24}$ & $\mathbf 2^{\prime \prime}$, $\mathbf 2^{\prime}$ & 2, 0 & 9, 3 & 3, 5 & 5, 3 & 7, 1 & 3, 3 & 0, 0 & 1, 3 \\
\bottomrule
\end{tabular}
\caption{Messenger fields used in our model. After integrating out these
fields we end up with the desired effective operators. For the sake of
brevity we do not list all mass terms in the text. The messenger pair
in every line has a mass term and there are no cross terms allowed.
\label{Tab:Messengers}}
\end{table}

In our model we consider non-renormalisable operators. In general
the contraction of the $SU(5)$ and $T^{\prime}$ indices may not be unique
which is nevertheless essential for our model. Our
predictions are based on the fact, that only a certain contraction
is allowed as we have, for example, indicated in eq.\ \eqref{eq:Wu} for the
$T^{\prime}$ indices. For the connection between the so-called
UV completion and predictivity of a model see also \cite{Varzielas:2010mp}.
Hence we have to specify the so-called messenger fields
which generate only the desired contractions in the operators after being
integrated out in a specific order.

The full list of messenger fields of our model is given in Tab.\
\ref{Tab:Messengers}. Every messenger pair in every line receives a
mass term in the superpotential, like, for example, $M_{\Sigma_{1}^a}  \Sigma_{1}^a
\bar{\Sigma}_{1}^a$. For the sake of brevity we do not write down all
mass terms, but it is important to note, that there are no mass terms
between messengers in different lines allowed. We assume all the
messenger masses to be above the scale of $T^{\prime}$ and $SU(5)$
breaking, which are closely related in our model as we will see in the
next section. Many messengers carry $SU(5)$
quantum numbers so that above the messenger scale, which we
denote by $\Lambda$ the gauge coupling becomes quickly
non-perturbative, so that we are not predictive above this scale.

\begin{figure}
\centering
\includegraphics[scale=0.39]{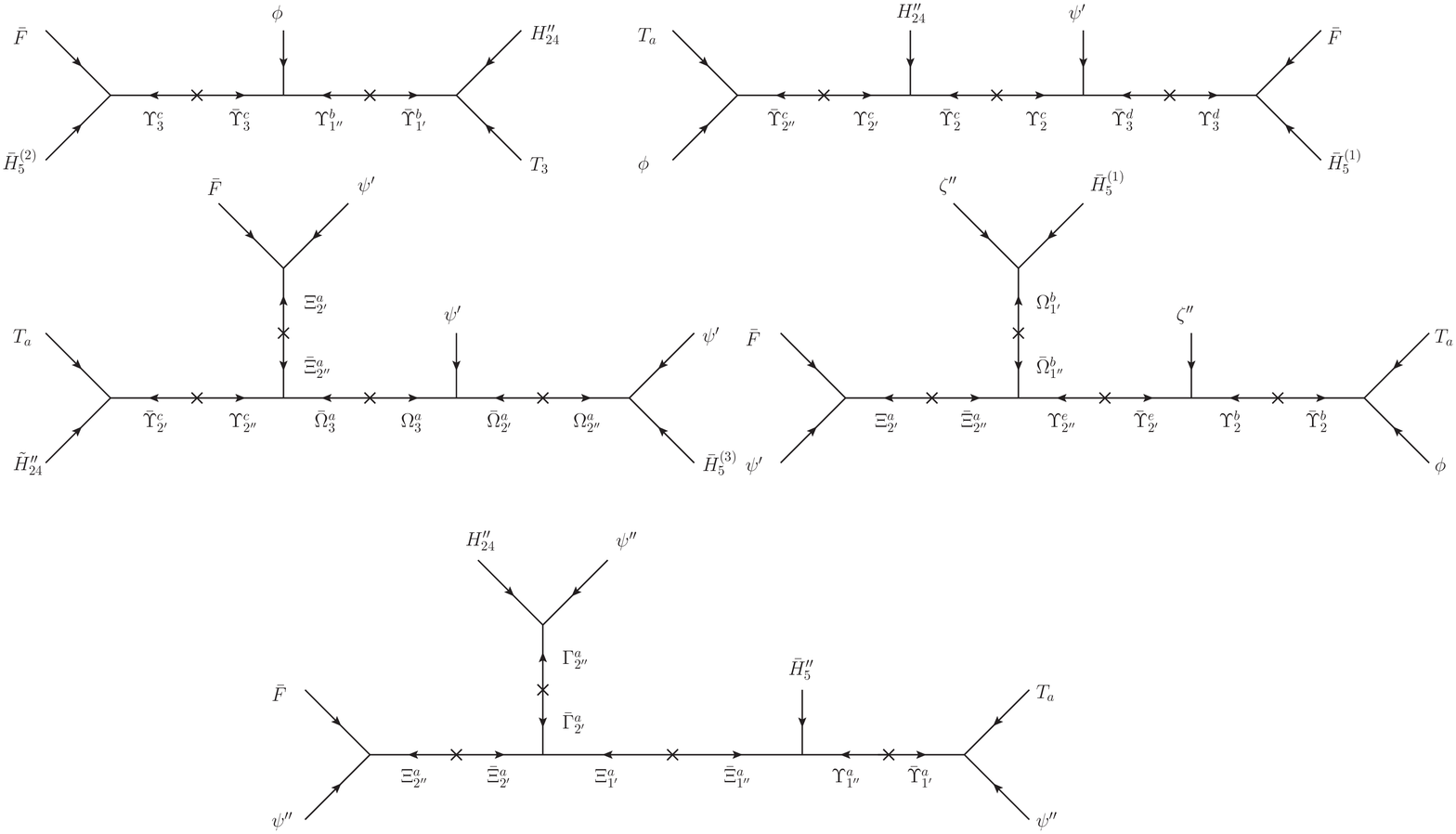}
\caption{
The supergraphs before integrating out the messengers for the down-type quark and charged lepton sector.
\label{Fig:DownMessenger} }
\end{figure}

\begin{figure}
\centering
\includegraphics[scale=0.47]{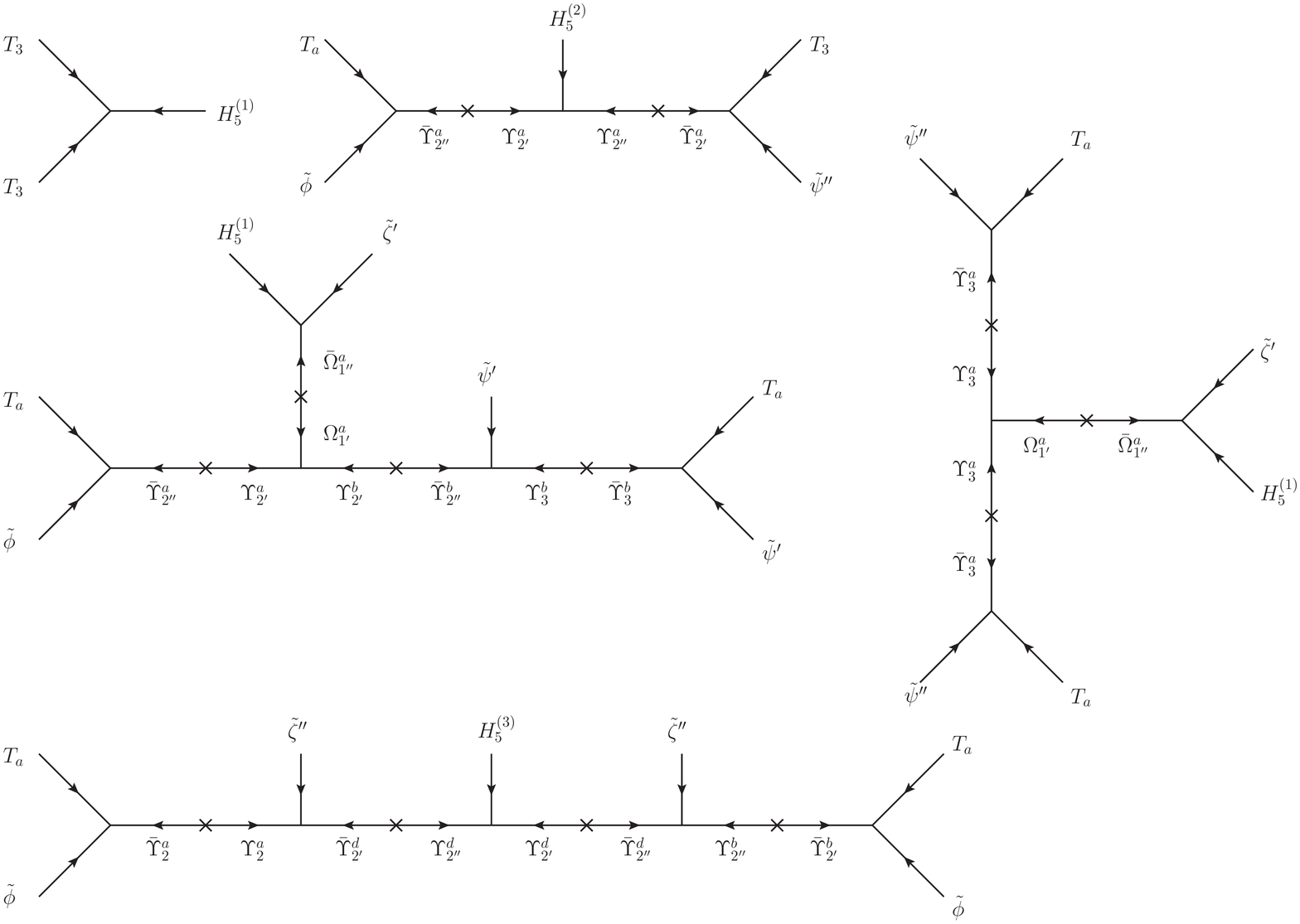}
\caption{ The supergraphs before integrating out the messengers for
the up-type quark sector. \label{Fig:UpMessenger} }
\end{figure}

\begin{figure}
\centering
\includegraphics[scale=0.6]{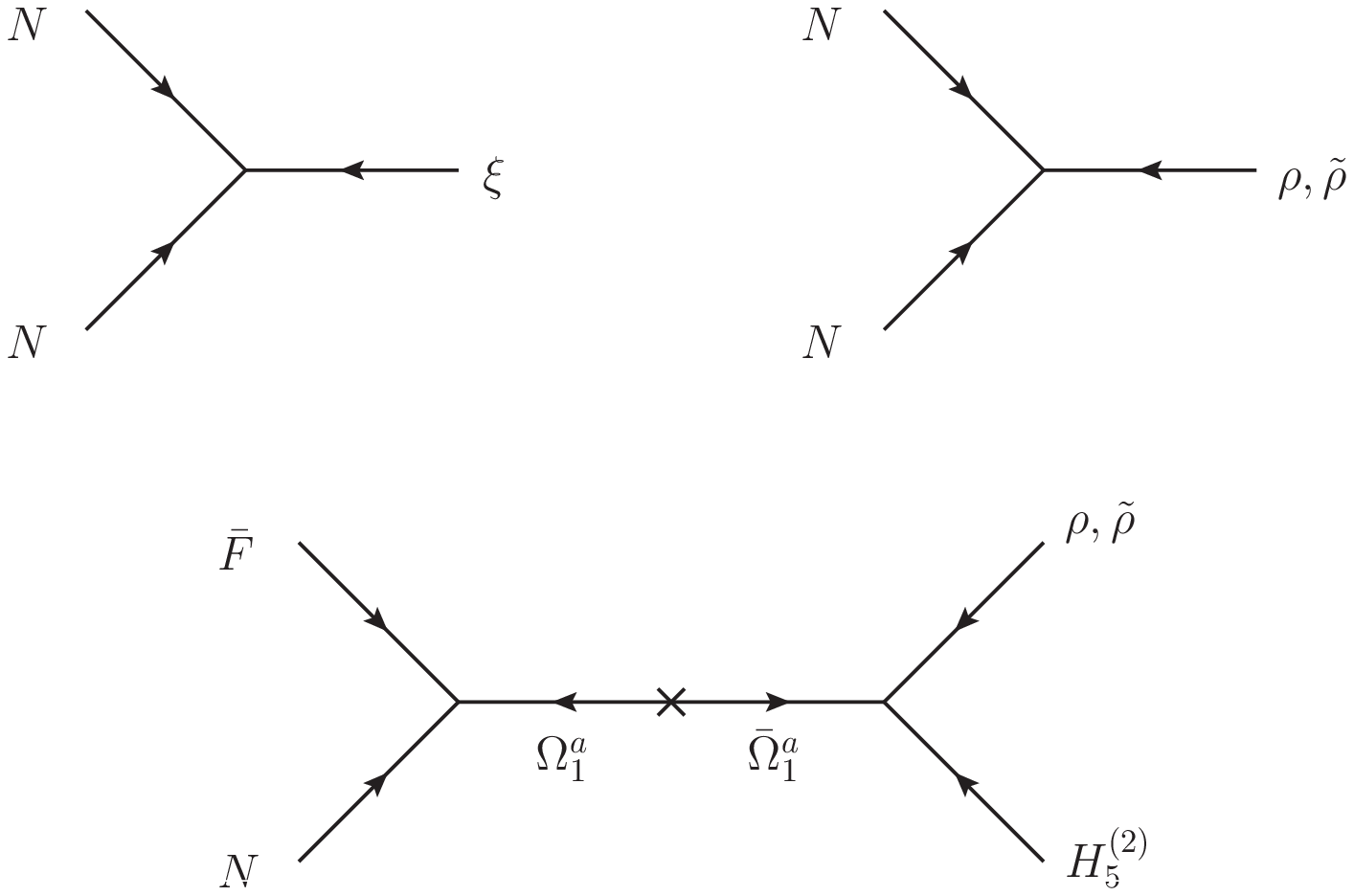}
\caption{
The supergraphs before integrating out the messengers for the neutrino sector.
\label{Fig:NeutrinoMessenger} }
\end{figure}

After this general remarks we now turn to the superpotential
which describes the couplings of the various fields to the
messengers. We start with the messengers coupling to the
matter,  Higgs and flavon fields. The supergraphs showing
these couplings are given in Figs.\ \ref{Fig:DownMessenger}-\ref{Fig:NeutrinoMessenger}.
From these diagrams one can
read off all the relevant contractions and couplings. Nevertheless,
we give now the renormalisable superpotential containing the
messenger fields.

Apart from the messenger mass terms (which we do not write
down explicitly) there are no terms with one or two fields
involving matter, Higgs and flavon fields.
For the down-type quark diagrams we find
(here and in this whole section we do not write down the couplings)
\begin{align}
\mathcal{W}^{\text{ren}}_d &=  \bar F \bar H_5^{(2)} \Upsilon_3^c
+ \phi \bar \Upsilon_3^c  \Upsilon_{1^{\prime \prime}}^b
+ H_{24}^{\prime \prime} T_3 \bar \Upsilon_{1^{\prime}}^b \\
&
+ T_a \phi \bar \Upsilon_{2^{\prime \prime}}^c
+ H_{24}^{\prime \prime} \bar \Upsilon_2^c  \Upsilon_{2^{\prime}}^c
+ \psi^{\prime} \Upsilon_2^c \bar \Upsilon_3^d
+ \bar F \bar H_5^{(1)} \Upsilon_3^d \\
&
+ T_a  \ti H_{24}^{\prime \prime} \bar  \Upsilon_{2^{\prime}}^c
+ \bar H_5^{(3)}  \psi^{\prime}  \Omega_{2^{\prime \prime}}^a
+ \bar \Omega_{2^{\prime}}^a \psi^{\prime} \Omega_3^a
+ \bar \Omega_3^a \Upsilon_{2^{\prime \prime}}^c \bar \Xi_{2^{\prime \prime}}^a \\
&
+ \bar F \psi^{\prime} \Xi_{2^{\prime}}^a
+ \bar \Xi_{2^{\prime \prime}}^a \Upsilon_{2^{\prime \prime}}^e \bar \Omega_{1^{\prime \prime}}^b
+ \zeta^{\prime \prime} \bar H_5^{(1)} \Omega_{1^{\prime}}^b
+ \zeta^{\prime \prime} \bar \Upsilon_{2^{\prime}}^e \Upsilon_2^b
+ T_a \phi \bar \Upsilon_2^b \\
&
+ \bar F \psi^{\prime \prime} \Xi_{2^{\prime \prime}}^a
+ \bar \Xi_{2^{\prime}}^a \Xi_{1^{\prime}}^a \bar \Gamma_{2^{\prime}}^a
+ H_{24}^{\prime \prime} \psi^{\prime \prime} \Gamma_{2^{\prime \prime}}^a
+ \bar H_5^{\prime \prime} \bar \Xi_{1^{\prime \prime}}^a \Upsilon_{1^{\prime \prime}}^a
+ \bar \Upsilon_{1^{\prime}}^a T_a \psi^{\prime \prime} \;,
\end{align}
for the up-type quarks
\begin{align}
\mathcal{W}^{\text{ren}}_u &= H_5^{(1)} T_3^2
+ T_a   \ti \phi  \bar \Upsilon_{2^{\prime \prime}}^a
+ H_5^{(2)}  \Upsilon_{2^{\prime}}^a  \Upsilon_{2^{\prime \prime}}^a
+ T_3  \ti \psi^{\prime \prime}  \bar\Upsilon_{2^{\prime}}^a \\
&
+ T_a \ti \psi^{\prime \prime} \bar \Upsilon_3^a
+ \ti \zeta^{\prime} H_5^{(1)} \bar \Omega_{1^{\prime \prime}}^a
+ \Omega_{1^{\prime}}^a \Upsilon_3^a \Upsilon_3^a \\
&
+ T_a \ti \phi \bar\Upsilon_{2^{\prime \prime}}^a
+ \Upsilon_{2^{\prime}}^a \Omega_{1^{\prime}}^a \Upsilon_{2^{\prime}}^b
+ H_5^{(1)} \ti \zeta^{\prime} \bar \Omega_{1^{\prime \prime}}^a
+ \ti \psi^{\prime} \bar\Upsilon_{2^{\prime \prime}}^b \Upsilon_3^b
+ \bar \Upsilon_3^b T_a \ti \psi^{\prime} \\
&
+ T_a \ti \phi \bar \Upsilon_2^a
+ \Upsilon_2^a \ti \zeta^{\prime \prime} \bar \Upsilon_{2^{\prime}}^d
+ H_5^{(3)}  \Upsilon_{2^{\prime \prime}}^d \Upsilon_{2^{\prime}}^d
+ \ti \zeta^{\prime \prime} \bar \Upsilon_{2^{\prime \prime}}^d \Upsilon_{2^{\prime \prime}}^b
+ T_a \ti \phi \bar \Upsilon_{2^{\prime}}^b \;,
\end{align}
and for the neutrino sector
\begin{align}
\mathcal{W}^{\text{ren}}_\nu &= N^2 \xi + N^2 \rho + N^2 \ti \rho
+ \bar F N \Omega_1^a +  H_5^{(2)} \rho  \bar \Omega_1^a +
 H_5^{(2)} \ti \rho  \bar \Omega_1^a \;.
\end{align}
There are five additional operators which generate dimension
eight or more operators in the matter sector, which we neglect.
For completeness we give them as well
\begin{equation}
\mathcal{W}^{\text{ren, matter}}_{d \geq 8} =
  \zeta^{\prime} \Upsilon_2^c \bar\Upsilon_{2^{\prime \prime}}^c
+ \bar\Upsilon_2^c \Upsilon_3^d \psi^{\prime \prime}
+ \psi^{\prime \prime}  \Omega_{2^{\prime \prime}}^a \bar\Omega_3^a
+ \bar \Gamma_{2^{\prime}}^a \bar \Upsilon_{2^{\prime \prime}}^c \Upsilon_3^d
+ \bar \Gamma_{2^{\prime}}^a \bar \Upsilon_2^b \Upsilon_3^d \;.
\end{equation}

\begin{figure}
\centering
\includegraphics[scale=0.8]{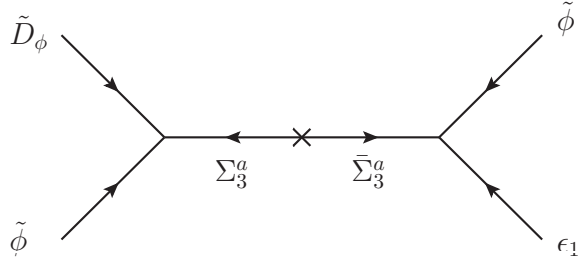}
\caption{One typical diagram for the messengers in the flavon sector.
  We consider only effective operators up to dimension four. For the sake
 of brevity we only show one diagram. The other ones are quite similar with
 the driving field on one side and the auxiliary $\epsilon$ fields on the
 other side.
\label{Fig:FlavonMessenger} }
\end{figure}

We turn now to the messengers, which give the
non-renormalisable terms in the flavon alignment
superpotential which we denote collectively with
$\Sigma$. In this sector all the supergraphs have the
structure as given in Fig.\ \ref{Fig:FlavonMessenger}.
(The role of the auxiliary $\epsilon$ fields is described
in the next section and their quantum numbers are given
in Tab.\ \ref{tab:FlavonAux}.)
For the sake of brevity we do not give all the diagrams
for the flavon sector, but all diagrams can easily be
derived from the renormalisable flavon superpotential.
We give here only the terms, where a messenger is
involved. The terms, where no messenger is involved
will be discussed in the next section, when we discuss
the superpotential responsible for the flavon alignment.
The superpotential involving the $\Sigma$ fields reads
\begin{align}
\mathcal{W}^{\text{ren}}_{\Sigma} &=
  D_{\xi}  \xi \Sigma_3^c
+ D_{\xi}  \rho \Sigma_3^c
+ D_{\xi}  \ti \rho \Sigma_3^c
+ \bar \Sigma_3^c \xi \epsilon_9
+ \ti D_{\phi} \ti \phi \Sigma_3^a
+ \bar \Sigma_3^a \ti \phi  \epsilon_1 \\
&
+ \ti D_{\phi}  \ti \phi  \Sigma_{1^{\prime \prime}}^a
+ \bar \Sigma_{1^{\prime}}^a \ti \zeta^{\prime \prime}  \epsilon_2
+ D_{\phi}  \phi  \Sigma_3^b
+ \bar \Sigma_3^b\phi  \epsilon_4
+ D_{\phi}  \phi  \Sigma_{1^{\prime \prime}}^b
+ \bar \Sigma_{1^{\prime}}^b \zeta^{\prime \prime}  \epsilon_5 \\
&
+ D_{\psi}  \psi^{\prime \prime}  \Sigma_{2^{\prime \prime}}^a
+ \bar \Sigma_{2^{\prime}}^a \psi^{\prime \prime}  \epsilon_6
+ D_{\psi}  \phi  \Sigma_{1^{\prime}}^a
+ S^{\prime \prime}_{\zeta} \epsilon_7 \Sigma_{1^{\prime}}^a
+ \bar \Sigma_{1^{\prime \prime}}^a \zeta^{\prime}  \epsilon_7
+ S_1  \psi^{\prime}  \Sigma_{2^{\prime \prime}}^b
+ \bar \Sigma_{2^{\prime}}^b \ti \psi^{\prime \prime} \epsilon_8 \\
&
+ S^{\prime}_2  \zeta^{\prime}  \Sigma_{1^{\prime}}^b
+ \bar \Sigma_{1^{\prime \prime}}^b \zeta^{\prime}  \epsilon_{12}
+ S^{\prime}_2  \ti \zeta^{\prime}  \Sigma_{1^{\prime}}^c
+ \bar \Sigma_{1^{\prime \prime}}^c \ti \zeta^{\prime}  \epsilon_{13}
+ S_{\epsilon_{12}}  \epsilon_{12}  \Sigma_1^a
+  \bar  \Sigma_1^a \epsilon_{12}^2 \\
&
+ S_{\epsilon_{13}} \epsilon_{13}  \Sigma_1^b
+ \bar  \Sigma_1^b \epsilon_{13}^2
+ \ti S_{24}^{\prime \prime}  \ti H^{\prime \prime}_{24}  \Sigma_{1^{\prime \prime}}^c
+  \bar  \Sigma_{1^{\prime}}^c \epsilon_{10}  \ti H^{\prime \prime}_{24}
+ \ti S_{24}^{\prime \prime} \xi \Sigma_3^d
+  \bar \Sigma_3^d   \epsilon_{11}  \xi \;.
\end{align}

Apart from these there are as well operators
which give dimension five operators in the
flavon alignment superpotential after integrating
out the messenger fields, which we
we will neglect. These operators are
\begin{align}
\mathcal{W}^{\text{ren, flavon}}_{d \geq 5} &=
   S^{\prime \prime}_\zeta \Sigma_1^a \Sigma_{1^{\prime}}^b
+ S^{\prime \prime}_{\ti \zeta} \Sigma_1^b \Sigma_{1^{\prime}}^c
+ S^{\prime \prime}_{\ti \zeta} \Sigma_{1^{\prime \prime}}^a \Sigma_{1^{\prime \prime}}^a
+ S^{\prime \prime}_{\zeta} \Sigma_{1^{\prime \prime}}^b  \Sigma_{1^{\prime \prime}}^b
+ S^{\prime \prime}_{\ti \zeta} \Sigma_3^a \Sigma_3^a
+ S^{\prime \prime}_{\zeta} \Sigma_3^b \Sigma_3^b \\
&
+ S_{24}^{\prime \prime} \Sigma_3^c \Sigma_3^c
+ S_{24}^{\prime \prime} \Sigma_3^d \Sigma_3^d
+ (S_{\epsilon_i} + S_{\xi} + S_{\rho}) \Sigma_3^c \Sigma_3^c
+ (S_{\epsilon_i} + S_{\xi} + S_{\rho}) \Sigma_3^d \Sigma_3^d \;.
\end{align}

Now we have discussed the messenger sector.
After integrating out the messengers from
the renormalisable superpotential we end up with the
effective operators which give us the
desired flavon vev alignments and structures
of the Yukawa matrices.

\section{Flavon Vacuum Alignment} \label{App:Alignment}

In this appendix we present the solution for our flavon vacuum
alignment. In the present model all the discussed results
crucially depend on the vev structure and on the fact that all
flavon vevs are real.

In the flavon potential two new kinds of fields are introduced.
First we have to add driving fields which are gauge singlets but
transform in a non trivial way under the family and shaping
symmetries and have a $U(1)_R$ charge of two. Minimising
the $F$-term equations of this fields will give us the correct
alignment (including phases) as one possible solution.
Second we introduce auxiliary fields $\epsilon_i$, $i=1,\ldots,13$,
which are singlets under $SU(5)$ and $T^{\prime}$, but they transform
in a non trivial way under the additional shaping symmetries.
They appear only in the flavon superpotential. Indeed, these fields
are introduced to compensate the charges of different operators, so
that they are related to each other in the $F$-term equations.
Note that we have to include for our alignment non-renormalisable
operators, where we restrict ourselves to operators with mass
dimension not higher than four in the superpotential.
The driving fields are listed in Table~\ref{tab:Driving} and the
auxiliary fields are listed in Table~\ref{tab:FlavonAux}.

%%%%%%%%%%%%%%%%%%%%%%%%%%%%%
\begin{table}
\centering
\begin{tabular}{c ccccc ccccc ccccc}
\toprule
           & $\tilde D_\phi$ & $\tilde S_{\psi}$ & $\tilde S^{\prime \prime}_{\zeta}$ & $\tilde S_{\zeta}$ & $D_\phi$ & $D_\psi$ & $S^{\prime \prime}_\zeta$ & $D_\xi$ & $S_\xi$ & $S_\rho$ & $S^{\prime \prime}_{24}$ & $\tilde S^{\prime \prime}_{24}$ & $S_1$ & $S^{\prime}_2$ & $S_{\epsilon_i}$  \\
\midrule
$T^\prime$ & $\mathbf{3}$ & $\mathbf{1}$ & $\mathbf{1}^{\prime \prime}$
           & $\mathbf{1}$ & $\mathbf{3}$ & $\mathbf{3}$ & $\mathbf{1}^{\prime \prime}$
           & $\mathbf{3}$ & $\mathbf{1}$ & $\mathbf{1}$ & $\mathbf{1}^{\prime \prime}$
           & $\mathbf{1}^{\prime \prime}$ & $\mathbf{1}$ & $\mathbf{1}^{\prime}$ & $\mathbf{1}$ \\
 $Z_{12}^u$ & 6& 0& 0& 0& 6& 0& 0& 6& 0& 0& 0& 6& 6& 4& 0\\
 $Z_{8}^d$ &4& 0& 0& 0& 0& 2& 4& 4& 0& 0& 0& 4& 5& 0& 0\\
 $Z_{8}^\nu $ & 4& 0& 0& 0& 4& 2& 4& 4& 0& 0& 0& 0& 2& 0& 0\\
 $Z_{8} $& 0& 4& 0& 4& 0& 2& 4& 4& 0& 0& 0& 0& 2& 0& 0\\
 $Z_{6} $ &1& 0& 4& 0& 3& 0& 0& 3& 0& 0& 0& 0& 5& 2& 0\\
 $Z_{6}^{\prime} $ &1& 0& 4& 0& 0& 3& 0& 3& 0& 0& 0& 3& 5& 2& 0\\
 $Z_{4} $& 0& 0& 0& 0& 0& 2& 0& 2& 0& 0& 0& 2& 1& 0& 0\\
\bottomrule
\end{tabular}
\caption{\label{tab:Driving} List of the driving fields from the
superpotential which give the desired vacuum alignment. All driving
fields are $SU(5)$ gauge singlets and charged under $U(1)_R$ with
charge $+2$.}
\end{table}
%%%%%%%%%%%%%%%%%%%%%%%%%%%%%%%%%%%%

%%%%%%%%%%%%%%%%%%%%%%%%%%%%%
\begin{table}
\centering
\begin{tabular}{c ccccc ccccc ccc}
\toprule &$\epsilon_1$ & $\epsilon_2$ & $\epsilon_3$ & $\epsilon_4$
& $\epsilon_5$ & $\epsilon_6$ & $\epsilon_7$ & $\epsilon_8$ &
$\epsilon_9$ & $\epsilon_{10}$ & $\epsilon_{11}$ & $\epsilon_{12}$& $\epsilon_{13}$ \\
\midrule
 $Z_{12}^u$& 6& 6& 0& 6& 6& 6& 6& 6& 6& 0& 6& 8& 8\\
 $Z_{8}^d$ &4& 4& 0& 4& 0& 4& 0& 4& 4& 0& 0& 0& 4\\
 $Z_{8}^\nu $ & 4& 0& 0& 0& 4& 0& 0& 4& 4& 0& 0& 0& 0\\
 $Z_{8} $ &0& 0& 4& 4& 0& 4& 0& 4& 4& 4& 0& 0& 0\\
 $Z_{6} $ & 3& 3& 0& 3& 3& 0& 0& 0& 3& 0& 0& 4& 0\\
 $Z_{6}^{\prime} $&  3& 3& 0& 3& 3& 0& 0& 0& 3& 0& 0& 4& 0\\
 $Z_{4} $& 0& 0& 0& 0& 2& 0& 2& 0& 2& 0& 2& 0& 0\\
\bottomrule
\end{tabular}
\caption{\label{tab:FlavonAux} List of the auxiliary flavon fields
that do not couple to the matter sector. The $\epsilon_i$ fields
are all $SU(5) \times T^{\prime}$ singlets and carry no $U(1)_R$
charge.}
\end{table}
%%%%%%%%%%%%%%%%%%%%%%%%%%%%%%%%%%%%

Before going into the more complicated details of the flavon vacuum
alignment we briefly discuss the ``alignment'' of the auxiliary flavons,
which is simply the question how to give them a real vev. For this
purpose we used the simple idea advocated in \cite{Antusch:2011sx},
which we can directly illustrate at the alignment for the $\epsilon$
fields itself. The superpotential for their alignment reads
\begin{align}
\mathcal{W}_{\epsilon} & = S_{\epsilon_i} \left( \epsilon_i^2  -
M_{\epsilon_i}^2\right) + S_{\epsilon_{j}} \left(
\frac{1}{\Lambda}\, \epsilon_{j}^3 - M_{\epsilon_{j}}^2 \right) \;,
\end{align}
where $i=1,\ldots,11$ and $j = 12,13$. Note that for the sake
of readibility we do not include any couplings. The driving fields
$S_{\epsilon_i}$ and $S_{\epsilon_j}$ are total singlets
so that terms like $S_{\epsilon} M_{\epsilon}^2$ are allowed.
The $F$-Term equations for the driving fields give, e.g.
\begin{equation}
 F_{S_{\epsilon_1}} = \epsilon_1^2  - M_{\epsilon_1}^2 = 0 \;.
\end{equation}
And since we assume that our fundamental theory is CP conserving
the mass $M_{\epsilon_1}$  is real (like the coupling parameters
which are not shown) and hence the vev of $\epsilon_1$
is real and non-vanishing. For $\epsilon_{12}$ and $\epsilon_{13}$
this has to be slightly modified. For them we find three possible
solutions, two of them complex and only one real. But we assume
that the real solution is picked up, which could be preferred by
higher order corrections, supergravity corrections or some low-energy
soft terms in the scalar potential. To discuss this corrections in
detail is beyond the scope of the current paper.

Note also that all the $S_{\epsilon_i}$ driving fields have the same
quantum numbers and hence can mix with each other. In other words each
of these driving fields could couple to each $\epsilon$ field.
We have chosen here the basis in which the superpotential has the
above structure, which makes the alignment clear (see also the
appendix of \cite{Antusch:2011sx}).

The same method can be applied to the real triplet and singlet
flavons of our model, after we have fixed their alignment by
some different kind of operators. But for the complex
doublets ($\mathbf{2}^{\prime}$, $\mathbf{2}^{\prime \prime}$)
and singlets ($\mathbf{1}^{\prime}$, $\mathbf{1}^{\prime \prime}$)
we have to use other relations, because the representation squared
cannot form a total singlet.

Before we come to this complex representations we discuss the
alignment for the flavons appearing in the neutrino sector ($\xi$,
$\rho$, $\ti \rho$) where this complication is absent \footnote{
The alignment for the triplets follows the discussion
in the seminal paper \cite{Altarelli:2005yx}.}.
The superpotential for these flavons reads
\begin{equation}
\mathcal{W}_{\xi, \rho, \ti \rho} =
    \frac{D_\xi}{\Lambda} \left(\xi^2 \epsilon_9 + \xi \rho \epsilon_9 + \xi \tilde\rho \epsilon_9 \right)
  + S_\xi \left(\xi^2 - M_{\xi}^2 \right)
  + S_\rho \left(\rho^2 + \tilde \rho^2 - M_{\rho}^2 \right) \;.
\end{equation}
The first thing to note here, is that we used the auxiliary
flavon $\epsilon_9$ in the first set of operators involving
the triplet driving field $D_{\xi}$. Since $\epsilon_9$
appears in all three operators, it drops out in the $F$-term
conditions, but nevertheless it is real and hence would just
modify the value of the vev without introducing any
phase. The $F$-term conditions are
\begin{align}
\frac{\partial \mathcal{W}_{\xi, \rho, \ti \rho} }{\partial  D_{\xi_1}} &= 2 \xi_1^2 - 2  \xi_2  \xi_3+\xi_1 (\rho+\ti\rho)=0 \;,\\
\frac{\partial \mathcal{W}_{\xi, \rho, \ti \rho} }{\partial  D_{\xi_2}} &= 2 \xi_2^2 - 2  \xi_1  \xi_3+\xi_3 (\rho+\ti\rho)=0 \;,\\
\frac{\partial \mathcal{W}_{\xi, \rho, \ti \rho} }{\partial  D_{\xi_3}} &= 2 \xi_3^2 - 2  \xi_2  \xi_1+\xi_2 (\rho+\ti\rho)=0 \;,\\
\frac{\partial \mathcal{W}_{\xi, \rho, \ti \rho} }{\partial  S_{\xi}}   &= \xi_1^2+2\xi_2 \xi_3- M_\xi^2=0 \;,\\
\frac{\partial \mathcal{W}_{\xi, \rho, \ti \rho} }{\partial  S_{\rho}}  &= \rho^2+\ti\rho^2-M_\rho^2=0 \;.
\end{align}
Besides the trivial solution $\xi_i =0$, $i=1,2,3$, we find
for the first three of these equations by cyclic permutations
in $\xi_i$ the desired solution for which $\xi_i = \xi_0
\neq 0$ if $\rho_0 = -\ti \rho_0$. The fact that the vevs are
non-vanishing and real can then be read off from the last two
equations for which we used the method from \cite{Antusch:2011sx}
discussed above.

Now we come to the most complicated part of the flavon alignment sector,
the flavons present in
the quark and charged lepton sectors. Although we have two different
set of flavons, one for the up-quark sector at the one hand and one
for the down-type quark and charged lepton sector at the other hand,
we cannot separate their alignments completely. In fact,
we found that the alignment is itself independent from each other
but the simplest solution which we found to make all vevs real
involves cross couplings between the two sectors.
The flavon superpotential reads
\begin{align} \mathcal{W}_f & =
    \frac{\ti D_{\phi}}{\Lambda} \left( \ti \phi \ti \phi \,\epsilon_1 + \ti \phi \ti \zeta^{\prime \prime} \epsilon_2 \right)
  + \ti S^{\prime\prime}_{\zeta} ( \ti \zeta^{\prime\prime} \ti \zeta^{\prime\prime} + \ti \phi \ti \phi - M_{\ti \zeta^{\prime}} \ti \zeta^{\prime})
  + \ti S_{\zeta} \, ( \ti \zeta^{\,\prime}\ti \zeta^{\,\prime \prime} - M_{\ti \zeta}\, \epsilon_{3}) \\
& + \ti S_{\psi} \left( \ti \psi^{\prime} \ti \psi^{\prime \prime}  - \ti \zeta^{\,\prime} \ti \zeta^{\,\prime \prime} \right)
  + \frac{D_{\phi}}{\Lambda} \left(\phi \phi\, \epsilon_4 +  \phi \,\zeta^{\,\prime \prime} \,\epsilon_5 \right)
  + \frac{D_{\psi}}{\Lambda} \left(\left( \psi^{\prime \prime}\right)^2 \epsilon_6  +  \phi\zeta^{\,\prime} \epsilon_7 \right) \\
& + S_{\psi} \left( \psi^{\prime} \psi^{\prime \prime} - M_{\psi}^2  \right)
  + S_{\zeta}^{\prime \prime} \left( \zeta^{\prime \prime} \zeta^{\prime \prime}  + \phi \phi - M_{\zeta^{\,\prime}} \zeta^{\prime} + \frac{\epsilon_7^2}{\Lambda} \zeta^{\prime} \right)\\
& + S_1 \left(\psi^{\,\prime}\, \ti \psi^{\,\prime\prime}\, \frac{ \epsilon_{8}}{\Lambda} - M_{S_1}^2\right)
  + S^{\prime}_2  \left( (\zeta^{\prime})^2 \epsilon_{12} - (\ti \zeta^{\prime})^2\epsilon_{13}\right)  \;,
\end{align}
where in the last equations the cross couplings between the two sectors
are written. We do not want to discuss here all the details of the
alignment in detail, instead we will only discuss the phases of the
vevs of the complex fields in a bit more detail. Nevertheless, we quote
all the $F$-term conditions, in which it is then quite easy to plug in
the flavon vevs from eqs.\ \eqref{eq:3FlavonAlignment}-\eqref{eq:1FlavonAlignment}
and see that they form a viable solution. The $F$-term conditions read for
the up sector
%%%%%%%%%%%%%%%%%%%%%%%%%%%%%%%%%%%%%%%%%%%%%%%%%%%%%%%
\begin{align}
 \frac{\partial \mathcal{W}_f }{\partial \ti D_{\phi_1}}&= \epsilon_1 (2 \ti \phi_1^2 - 2 \ti \phi_2 \ti \phi_3) + \epsilon_2 \ti \phi_2 \ti \zeta^{\prime\prime}=0 \;,\\
 \frac{\partial \mathcal{W}_f }{\partial \ti D_{\phi_2}}&= \epsilon_1 (2 \ti \phi_2^2 - 2 \ti \phi_1 \ti \phi_3) + \epsilon_2 \ti \phi_1 \ti \zeta^{\prime\prime}=0 \;,\\
 \frac{\partial \mathcal{W}_f }{\partial \ti D_{\phi_3}}&= \epsilon_1 (2 \ti \phi_3^2 - 2 \ti \phi_1 \ti \phi_2) + \epsilon_2 \ti \phi_3 \ti \zeta^{\prime\prime}=0 \;,\\
 \frac{\partial \mathcal{W}_f }{\partial \ti S^{\prime\prime}_{\zeta}} &= (\ti \zeta^{\prime\prime})^2 - M_{\ti\zeta^{\prime}}\ti\zeta^{\prime} + \ti \phi_3^2 + 2 \ti \phi_1 \ti \phi_2 =0 \label{eq:TZetaAlignment1} \;,\\
 \frac{\partial \mathcal{W}_f }{\partial \ti S_{\zeta}} &= \ti \zeta^{\prime}\ti \zeta^{\prime\prime} - M_{\ti\zeta}\epsilon_3=0 \label{eq:TZetaAlignment2} \;,\\
 \frac{\partial \mathcal{W}_f }{\partial \ti S_{\psi}} &= \ti \psi_1^{\prime} \ti \psi_2^{\prime\prime}- \ti \psi_2^{\prime} \ti \psi_1^{\prime\prime} - \ti \zeta^{\prime}\ti \zeta^{\prime\prime}=0 \;,
\end{align}
%%%%%%%%%%%%%%%%%%%%%%%%%%%%%%%%%%%%%%%%%%%%%%%%%%%%%%%%%%%%%%%%
for the down sector
%%%%%%%%%%%%%%%%%%%%%%%%%%%%%%%%%%%%%%%%%%%%%%%%%%%%%%%%%%%%%%%%
\begin{align}
 \frac{\partial \mathcal{W}_f }{\partial  D_{\phi_1}}&= \epsilon_4 (2 \phi_1^2 - 2  \phi_2  \phi_3) + \epsilon_5  \phi_2 \zeta^{\,\prime\prime}=0 \;,\\
 \frac{\partial \mathcal{W}_f }{\partial  D_{\phi_2}}&= \epsilon_4 (2 \phi_2^2 - 2  \phi_1  \phi_3) + \epsilon_5  \phi_1 \zeta^{\,\prime\prime}=0 \;,\\
 \frac{\partial \mathcal{W}_f }{\partial  D_{\phi_3}}&= \epsilon_4 (2 \phi_3^2 - 2  \phi_1  \phi_2) + \epsilon_5  \phi_3 \zeta^{\,\prime\prime}=0 \;,\\
 \frac{\partial \mathcal{W}_f }{\partial  D_{\psi_1}}&= \epsilon_6 ((\psi_2^{\,\prime\prime})^2+ \epsilon_7 \phi_3 \zeta^{\,\prime}=0 \;,\\
 \frac{\partial \mathcal{W}_f }{\partial  D_{\psi_2}}&= \ci \epsilon_6 ((\psi_1^{\,\prime\prime})^2+ \epsilon_7 \phi_2 \zeta^{\,\prime}=0 \;,\\
 \frac{\partial \mathcal{W}_f }{\partial  D_{\psi_3}}&= (1-\ci)\epsilon_6 \psi_1^{\,\prime\prime}\psi_2^{\,\prime\prime} + \epsilon_7 \phi_1  \zeta^{\,\prime}=0 \;,\\
 \frac{\partial \mathcal{W}_f }{\partial  S_{\psi}}&= \psi_1^{\prime}  \psi_2^{\,\prime\prime}-  \psi_2^{\prime} \psi_1^{\,\prime\prime} -  M_{\psi}^2=0 \;,\\
 \frac{\partial \mathcal{W}_f }{\partial S_{\zeta}^{\prime\prime}} &= ( \zeta^{\,\prime\prime})^2 - \left( M_{\zeta^{\prime}} + \frac{\epsilon_7^2}{\Lambda} \right) \zeta^{\prime} =0 \;, \label{Eq:ZetaAlignment1}
\end{align}
%%%%%%%%%%%%%%%%%%%%%%%%%%%%%%%%%%%%%%%%%%%%%%%%%%%%%%%%%%%%%%%%
and for the cross couplings between the two sectors
%%%%%%%%%%%%%%%%%%%%%%%%%%%%%%%%%%%%%%%%%%%%%%%%%%%%%%%%%%%%%%%%
\begin{align}
 \frac{\partial \mathcal{W}_f }{\partial S_1}&= \left(\psi_1^{\prime} \ti \psi_2^{\,\prime\prime}-  \psi_2^{\prime} \ti \psi_1^{\,\prime\prime} \right) \frac{\epsilon_{8}}{\Lambda} -M_{S_1}^2 = 0 \;,\\
 \frac{\partial \mathcal{W}_f }{\partial S^{\prime}_2}&=(\zeta^{\,\prime})^2\epsilon_{12}-(\ti\zeta^{\,\prime})^2 \epsilon_{13} = 0 \;. \label{Eq:ZetaAlignment2}
\end{align}
So how do we make the vevs of the complex representations real?
Exemplary we discuss the complex singlets $\ti \zeta
^{\prime \prime}$, $\ti \zeta^{\prime}$, $\zeta^{\prime \prime}$
and $\zeta^{\prime}$.
From eqs.\ \eqref{eq:TZetaAlignment1} and \eqref{eq:TZetaAlignment2}
we find a polynomial in $\ti \zeta^{\prime \prime}$
\begin{equation}
 (\ti \zeta^{\prime \prime})^3 + \ti \zeta^{\prime \prime} (\ti \phi_3^2 + 2 \ti \phi_1 \ti \phi_2)  - M_{\ti\zeta^{\prime}} M_{\ti\zeta}  \epsilon_3 = 0 \;,
\end{equation}
which has a real solution (at least for a
certain choice of parameters and plugging in the
real vev of $\ti \phi$) which we pick here.
Then we know that $(\ti \zeta^{\prime \prime})^3$ is real,
while $\ti \zeta^{\prime}$ has the opposite phase of $\ti
\zeta^{\prime \prime}$ so it is real as well.
From eq.\ \eqref{Eq:ZetaAlignment2} we
then find $\zeta^{\prime}$ to be real and from
eq.\ \eqref{Eq:ZetaAlignment1} we obtain $\zeta^{\prime
\prime}$ to be real and all the singlet vevs are real.
For the doublets a similar mechanism applies.

The last alignment we want to discuss here is strictly speaking
not an alignment. But since we have used adjoints of $SU(5)$
in our operators to get the desired Yukawa coupling relations
between the charged leptons and the down-type quarks we add here
a mechanism which generates the vev of these adjoints and also
show explicitly that they are real. For the fields $H_{24}^{\prime
\prime}$ and $\ti H_{24}^{\prime \prime}$ we can write down the
following superpotential using the two driving fields
$S_{24}^{\prime \prime}$ and $\tilde S_{24}^{\prime \prime}$
\begin{equation}
\mathcal{W}_{24} =
    S_{24}^{\prime \prime} \left( H_{24}^{\prime \prime} H_{24}^{\prime \prime} - \xi^2 \right)
  + \frac{\ti S_{24}^{\prime \prime}}{\Lambda} \left( \tilde H_{24}^{\prime \prime} \tilde H_{24}^{\prime \prime} \epsilon_{10} - \xi^2 \epsilon_{11} \right) \;.
\end{equation}
We see that the vev of $\xi$ triggers a vev
for the two adjoint fields and even more these
two vevs are directly related to the $T^{\prime}$
symmetry breaking scale. That means that in our
model the GUT scale and the scale of $T^{\prime}$
coincide (up to some order one coefficients).
In principle, we can again choose here between two
different vevs for the adjoints. One pointing into
the SM direction and the other one pointing into the
$SU(4) \times U(1)$ direction and we assume the first
option to be realised. We also note here, that the
solution of the Doublet-Triplet-Splitting problem
and hence the construction of the whole Higgs sector
is clearly beyond the scope of this paper.

\end{document}